%% file: ms.tex
\pdfoutput=1

\newif\ifreview
\reviewfalse

\newif\ifnfs
\nfsfalse

\newif\iftext
\textfalse

\newif\ifarxiv
\arxivtrue

\ifreview
	\iftext
		\documentclass[12pt,onecolumn,peerreview,draft]{IEEEtran}
	\else
		\documentclass[12pt,onecolumn,peerreview,draftcls]{IEEEtran}
	\fi
\else
	\iftext
		\documentclass[10pt,twocolumn,journal]{IEEEtran}
	\else
		\documentclass[10pt,twocolumn,journal,final]{IEEEtran}
	\fi
\fi

\input{setup}

\begin{document}

\input{preamble}

\input{introduction}

\input{background}

\input{methods}

\input{results}

\input{conclusion}

\input{postamble}

\end{document}

%% file: setup.tex
\usepackage{cite}
\usepackage{etoolbox}
\AtBeginEnvironment{quote}{\small}
\usepackage{amsmath}
\usepackage{amsthm}
\usepackage{booktabs}
\usepackage[font=small]{caption}
\usepackage[flushleft]{threeparttable}
\usepackage{breqn}
\usepackage{ftnright}
\usepackage{hyperref}
\usepackage{mathrsfs}
\DeclareMathAlphabet{\mathpzc}{OT1}{pzc}{m}{it}
\usepackage{subcaption}

\DeclareMathOperator*{\argmax}{arg\,max}
\DeclareMathOperator*{\argmin}{arg\,min}

\ifnfs
  
\else
  
\fi

\ifCLASSINFOpdf
  \iftext
    \usepackage[pdftex,draft]{graphicx}
  \else
    \usepackage[pdftex]{graphicx}
  \fi
  \ifnfs
    \graphicspath{{./cephfigures/}{./figures/}{./diagrams/}}
  \else
    \graphicspath{{./figures/}{./diagrams/}}
  \fi
\else
\fi

\usepackage[ruled,vlined]{algorithm2e}

%% file: preamble.tex
%
%
\title{Remote Sensor Design for Visual Recognition with Convolutional Neural Networks}
%
%
%

\ifreview
\else
	\author{Lucas~Jaffe,
	        Michael~Zelinski,
	        and~Wesam~Sakla
	\ifarxiv
	\thanks{\copyright 2019 IEEE.  Personal use of this material is permitted.  Permission from IEEE must be obtained for all other uses, in any current or future media, including reprinting/republishing this material for advertising or promotional purposes, creating new collective works, for resale or redistribution to servers or lists, or reuse of any copyrighted component of this work in other works.}
	\else
	\fi
	\thanks{L. Jaffe, M. Zelinski, and W. Sakla are research scientists at Lawrence Livermore
	National Laboratory, Livermore, CA 94550 USA. }
	\thanks{This work was performed under the auspices of the U.S. Department of Energy by Lawrence Livermore National Laboratory under Contract DE-AC52-07NA27344.}
	\thanks{This document was prepared as an account of work sponsored by an agency of the United States government. Neither the United States government nor Lawrence Livermore National Security, LLC, nor any of their employees makes any warranty, expressed or implied, or assumes any legal liability or responsibility for the accuracy, completeness, or usefulness of any information, apparatus, product, or process disclosed, or represents that its use would not infringe privately owned rights. Reference herein to any specific commercial product, process, or service by trade name, trademark, manufacturer, or otherwise does not necessarily constitute or imply its endorsement, recommendation, or favoring by the United States government or Lawrence Livermore National Security, LLC. The views and opinions of authors expressed herein do not necessarily state or reflect those of the United States government or Lawrence Livermore National Security, LLC, and shall not be used for advertising or product endorsement purposes.}
	}
\fi
%
%

\ifarxiv
\markboth{\textcopyright 2019 IEEE. ACCEPTED FOR PUBLICATION IN IEEE TRANSACTIONS ON GEOSCIENCE AND REMOTE SENSING.}%
\markboth{}
\else
\fi
%

\IEEEoverridecommandlockouts
\IEEEpubid{
\begin{minipage}{\textwidth}\ \\[12pt]
  LLNL-JRNL-760588                                                                                                                                                                                                                                                                                                                                                                                                                                                                                                                                                                                                                                                                                                                                                                                                                               
\end{minipage}} 

\maketitle

\begin{abstract}
	While deep learning technologies for computer vision have developed rapidly since 2012, modeling of remote sensing systems has remained focused around human vision. In particular, remote sensing systems are usually constructed to optimize sensing cost-quality trade-offs with respect to human image interpretability. While some recent studies have explored remote sensing system design as a function of simple computer vision algorithm performance, there has been little work relating this design to the state-of-the-art in computer vision: deep learning with convolutional neural networks. We develop experimental systems to conduct this analysis, showing results with modern deep learning algorithms and recent overhead image data. Our results are compared to standard image quality measurements based on human visual perception, and we conclude not only that machine and human interpretability differ significantly, but that computer vision performance is largely self-consistent across a range of disparate conditions. This research is presented as a cornerstone for a new generation of sensor design systems which focus on computer algorithm performance instead of human visual perception.
\end{abstract}


%
\IEEEpeerreviewmaketitle


%% file: introduction.tex
\section{Introduction}

\IEEEPARstart{I}{n} the past decade, overhead image data collection rates have expanded to the petabyte scale \cite{high_data_volume, moody2016building, xia2017exploiting}. This imagery is used for a wide variety of applications including disaster relief \cite{application_disaster_relief}, environmental monitoring \cite{application_environmental}, and Intelligence, Surveillance, and Reconnaissance (ISR) \cite{application_isr}. Given the scale of this data and the need for fast processing, human analysis alone is no longer a tractable solution to these problems. 

Correspondingly, algorithmic processing for this imagery has improved rapidly, with a recent emphasis on the use of convolutional neural networks (CNNs). CNNs, originally proved significant by Yann LeCun in the late 1990's for solving simple image recognition problems \cite{lecun1999object}, rose to prominence in 2012 after a breakthrough many-layer (deep) architecture was introduced by Alex Krizhevsky \cite{krizhevsky2012imagenet}. The use of learned feature representations from CNNs quickly replaced a host of hand-crafted image features, including Histogram of Oriented Gradients (HOG) \cite{dalal2005histograms}, Local Binary Patterns (LBP) \cite{ojala2002multiresolution}, and Haar-like features \cite{viola2001rapid}.

In nearly every major visual recognition contest since 2012, all top solutions have used CNNs \cite{ILSVRC15, minetto2018hydra, zeiler2014visualizing}. Likewise, the top performing solutions on nearly all major visual recognition datasets employ CNNs\footnote{Best results on standard datasets:\\\url{http://rodrigob.github.io/are_we_there_yet/build/}} \cite{wan2013regularization, springenberg2014striving, snoek2015scalable, yang2015deep, lee2016generalizing}. A variety of robust neural network software frameworks have been developed in recent years to capitalize on the need for high-level network architecture design and deployment capability. These frameworks, including TensorFlow, Caffe, and PyTorch have enabled a large and active community working towards the development of neural networks for visual recognition.

Given the importance of visual recognition in overhead imagery, the criticality of using machines for this task, and the dominance of CNNs as a vehicle for doing so, it follows that we want to acquire imagery which is well-suited to visual recognition with CNNs. Consequently, the sensing systems which acquire this imagery must be designed to gather imagery optimal for visual recognition with CNNs. 

Broadly, we pose the following problem:
given a dataset of images $\mathcal{D}$ collected from a sensor with parameters $P$, find the set of values for $p \in P$ which optimize the performance of some image recognition model $m \in M$ on dataset $\mathcal{D}$ according to an objective function $L$ of desired metrics. Note that this $L$ would include the monetary cost of the sensor in a real-world setting. The problem to be solved e.g., classification, detection, or retrieval, is treated as a part of the model.
\begin{equation}
\label{principle_equation}
\argmin_{p \in P}{L(P; m, \mathcal{D})}
\end{equation}

This work describes how this optimization problem can be constructed using recent image data and state-of-the-art visual recognition methods, and we present results for several configurations of the problem. Importantly, the models $m \in M$ that we choose are CNN variants and image quality algorithms, but we emphasize that the method is general and can be applied for all $m \in M$, including the Human Visual System (HVS). We compare metrics of human recognition with metrics of machine recognition, and demonstrate an important pattern in the relationship between these metrics and the observed sensor parameters.

\begin{figure*}[!t]
\centering
\includegraphics[width=\textwidth,draft=false]{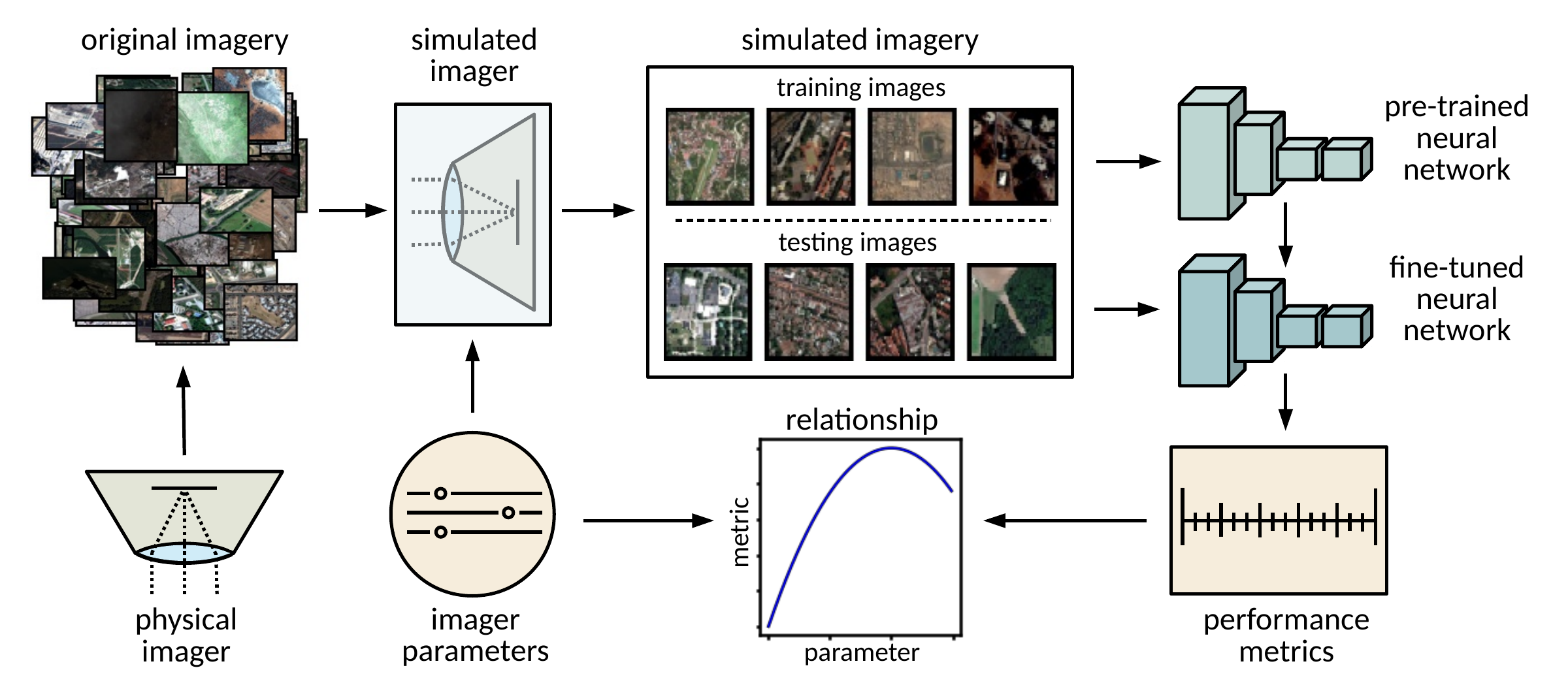} 
\caption{A diagram of our approach. We take existing imagery captured from some image system, then regenerate that imagery as though it were captured by an image system with parameters of our choosing. This regenerated imagery is partitioned into standard training and testing sets, then used to train a CNN classifier model. Results on the testing imagery are collected, and the process is repeated for many different parameter configurations in order to observe relationships between imager parameters and performance metrics.}
\label{fig:system}
\end{figure*}

A diagram of our approach is presented in Figure \ref{fig:system}. The variables and parameters of Equation \ref{principle_equation} correspond directly to the elements of this diagram.

In Section \ref{sec:background}, background is provided for relevant concepts in image utility and sensor modeling. In Section \ref{sec:methods}, the sensor simulator and experimental process are described. In Section \ref{sec:results}, the results of several experiments are explained and analyzed. Conclusions are presented in Section \ref{sec:conclusion}.

%% file: background.tex
\section{Background}
\label{sec:background}
\subsection{Image Quality Assessment}
The problem of \textit{image quality assessment} (IQA) is often divided into two categories: \textit{full-reference} IQA, in which an original, undistorted image is available to compare against the target image, and \textit{no-reference} IQA, in which no such original image is available for comparison \cite{wang2004image}. Many algorithms have been developed in each category, which produce a single scalar ``score'' value as output.

Metrics including Mean-Squared Error (MSE) and Peak Signal-to-Noise Ratio (PSNR) are often used for full-reference IQA, commonly to evaluate the effect of a compression algorithm such as JPEG. In  \cite{wang2004image}, Wang et al. argue that these basic metrics do not account for structural similarity perceived by the Human Visual System (HVS). They propose an index for structural similarity, SSIM, which computes a summary of window comparisons between two images, using means and standard deviations of pixels in the windows.

Metrics developed for no-reference IQA include the Blind/Referenceless Image Spatial Quality Evaluator (BRISQUE) \cite{mittal2012no} and the Natural Image Quality Evaluator (NIQE) \cite{mittal2013making}. The BRISQUE algorithm is considered \textit{opinion-aware}, because it is trained on human-annotated image quality evaluations. BRISQUE produces a measure of quality based on ``locally normalized luminance coefficients'', which the authors claim is sufficient to quantify both image naturalness and image distortion. The luminance features are used to train a SVM regressor which produces the final quality score. In contrast, the NIQE algorithm is trained on a corpus of natural images which are \textit{not} annotated for quality or distortion. NIQE computes \textit{natural scene statistic} (NSS) features on this corpus, which are similar to the BRISQUE luminance features. The final NIQE score is computed by comparing a multivariate Gaussian (MVG) fit of the corpus features to the MVG fit of the target image features.

By comparing these IQA metrics with human recognition and CNN recognition, we can assess whether they are similar to either of the two, and if they could be used to model CNN performance for different sensor systems.

\subsection{Image Utility}
We contrast the concept of image quality with \textit{image utility}, which explicitly concerns determining the value of an image for performing some recognition task. The National Imagery Interpretability Rating Scale (NIIRS) was released by the IRARS committee in 1974 for enumerating human interpretability of imagery, usually collected from overhead platforms \cite{leachtenauer1997general}. NIIRS defines a 0-9 rating scale for imagery, with 0 corresponding to complete uninterpretability (due to artifacts, blurring, noise, or occlusion), and 9 corresponding to interpretability of objects at the centimeter scale.

While ``image quality'' is ill-defined, image utility in the sense defined by NIIRS can be measured, tested, and optimized for. Equations have been developed to fit human-annotated NIIRS data, in order to relate common properties of imagery and corresponding collection systems to NIIRS. The General Image Quality Equation (GIQE) is the most well-known of these equations \cite{leachtenauer1997general}. We will discuss version 5 of the GIQE, the most recent, shown in Equation \ref{eq:giqe5}. Values for the constants in this equation are available in \cite{harrington2015general}.

\begin{dmath}
\label{eq:giqe5}
\text{NIIRS} = A_0 + A_1\log_{10}{\text{GSD}} + A_2 \left[ 1 - \exp{\frac{A_3}{\text{SNR}}} \right] \log_{10}{\text{RER}} + A_4\log_{10}{(\text{RER})^4} + \frac{A_5}{\text{SNR}}
\end{dmath}

The GIQE5 assumes an analyst will apply their own image enhancements, known as the \textit{softcopy} scenario, and that no generic enhancements are applied prior to viewing \cite{harrington2015general}. We consider the experiments in this work to be analagous to the assumptions of the GIQE5, because no enhancements are applied to the imagery, and the model learns to apply whatever corrections are beneficial to recognition in the imagery.

The GIQE5 has three variables, here explained: Ground Sample Distance (GSD) is the width of a pixel projected from image space onto the ground, given in units of distance per pixel. Signal-to-Noise Ratio (SNR) is the ratio of true signal to noise, mainly consisting of photon noise, dark noise, and read noise. Relative Edge Response (RER) measures sharpness of edges in the imagery \cite{leachtenauer1997general}. 

Of importance to this work are the concepts of \textit{f-number} (FN) and \textit{optical Q}, defined as:
\begin{equation}
\text{FN} = \frac{f}{D}
\quad\mathrm{and}\quad 
\text{Q} = \frac{\lambda \text{FN}}{p}
\end{equation}
where $f$ is the \textit{focal length} of the optic, $D$ is the \textit{diameter} of the optic aperture, $p$ is the \textit{pixel pitch} of the detector array, and $\lambda$ is the shortest \textit{wavelength} of light measured. In \cite{fiete1999image}, Fiete shows that optical Q is directly related to image properties associated with quality. For instance, reducing Q increases the SNR and improves image sharpness, but can also lead to aliasing below a certain level.

An important critique of our comparison to GIQE5 is that the equation assumes Q is between 1.0 and 2.0, whereas we mainly study the regime of Q $<$ 1.0. Fiete notes that reducing Q,  ``much below 1.0 can cause objectionable aliasing artifacts'' \cite{fiete1999image}. We leave a study of the Q regime between 1.0 and 2.0 to future work.

NIIRS and the GIQE are useful for tasking of existing imaging systems \cite{irvine1997national}, analyzing capabilities of existing imaging systems \cite{wong2014predicting}, or for the design of new imaging systems \cite{cota2009use}. Consider an image system design scenario: a remote sensing system engineer might select an aperture diameter for the system lens by examining its impact on GSD, SNR, the subsequent NIIRS value, and corresponding tradeoffs with system cost. While increasing optic aperture diameter can give lower GSD and higher SNR, and therefore higher NIIRS, the cost to build and launch the resulting system could create an upper bound for aperture diameter. If we relate this example to Equation \ref{principle_equation}, it can be formulated as:

\begin{equation}
\label{principle_equation_ex1}
\argmin_{D}{L(D; P, \mathcal{D})}
\end{equation}
with
\begin{equation}
L = \alpha \cdot \text{NIIRS}(D; P, \mathcal{D}) + \beta \cdot \text{cost}(D; P)
\end{equation}
where $D$ is the aperture diameter of the system, cost is the monetary cost of the system, NIIRS corresponds to predicted NIIRS for the system, and $\mathcal{D}$ more abstractly corresponds to imagery used to produce a model for NIIRS (possibly the GIQE). The objective function we optimize, $L$, could be a weighted combination of NIIRS and cost, weighted by some parameters $\alpha$ and $\beta$ respectively. All sensing system parameters except $D$ comprise $P$. The objective is parametrized by $P$ and $\mathcal{D}$. Note that in this case, we are optimizing for the HVS by optimizing NIIRS.

While NIIRS and the GIQE have been thoroughly studied, there has been no published evidence to date that they are useful for describing image interpretability for modern computer vision algorithms. More generally, we do not know if human and computer recognition of imagery are similar at all. The experiments detailed in Section \ref{sec:results} aim to explore this question by comparing machine learning metrics with the GIQE for the same system configuration.

Other approaches have alternatively framed the image interpretability problem in terms of \textit{target acquisition performance}. In \cite{vollmerhausen2004targeting}, Vollmerhausen and Jacobs give a detailed history of target acquisition performance, and introduce the Target Task Performance (TTP) metric. This metric, in addition to its predecessor, the Johnson criteria \cite{johnson1985analysis}, are Modulation Transfer Function (MTF)-based metrics \cite{gallimore1991review}. Since NIIRS and the GIQE have been more widely-adopted in practice, we focus our comparison against them.

\subsection{CNNs for Overhead Imagery}
CNNs have rapidly risen to the forefront of algorithms for visual recognition in overhead imagery \cite{zhu2017deep}. Most commonly, CNNs have been used to tackle the classification, retrieval, detection, and segmentation problems. In \cite{chen2016deep, mou2017deep},  different deep CNN architectures are used for image classification in overhead imagery. In \cite{xia2017exploiting}, a systematic investigation of image retrieval with deep learning is conducted for remote sensing imagery. In \cite{long2017accurate}, a CNN-based model is used for object localization (detection) in overhead imagery. In \cite{maggiori2017convolutional}, a fully convolutional architecture is developed for pixelwise classification (semantic segmentation) of satellite imagery.

In addition, large public datasets of overhead imagery are being curated and released at an increasing rate. The SpaceNet dataset, originally released in 2016, comprises hundreds of thousands of polygon labelings for buildings in DigitalGlobe satellite imagery, suited to the segmentation problem \cite{spacenet}. In the xView dataset, one-million objects across 60 classes are annotated with bounding boxes for the detection problem \cite{xview}. In the Functional Map of the World (fMoW) dataset, one-million DigitalGlobe images across 63 classes are annotated with rectangular bounding boxes for the classification problem \cite{fmow}.

This increase in public data volume, combined with innovations in GPU hardware and CNN architectures, have led to a boom in the use of CNNs for overhead imagery. 

\subsection{Sensor Modeling}
A number of tools have been developed for remote sensor modeling, including for image transformation and GIQE computation. LeMaster et al.\cite{lemaster2017pybsm} have developed the ``Python Based Sensor Model'' (pyBSM), which implements the ``ERIM Image Based Sensor Model'' (IBSM) \cite{eismann1996utility}. All GIQE values presented in Section \ref{sec:results} were computed with pyBSM.

Similarly, the Night Vision Integrated Performance Model (NV-IPM) is a software package developed by CERDEC's Night Vision and Electronic Sensors Directorate for modeling image system design\footnote{NV-IPM Web Page:\\\url{https://www.cerdec.army.mil/inside_cerdec/nvesd/integrated_performance_model/}}.

A description of our remote sensor modeling approach is presented in Section \ref{sec:methods}.

\subsection{CV-Based Modeling}
While remote sensor system design is often based on human interpretability through NIIRS or the TTP, there are studies which have measured performance using IQA algorithms, or simple visual recognition algorithms.

In \cite{fanning2012metrics}, Fanning develops a custom MTF-based image generation tool, and uses it to conduct a comparitive study based on the SSIM metric. 

In \cite{lemaster2017pybsm}, LeMaster et al. demonstrate how the face detection problem could be used to impact sensor design by transforming imagery with pyBSM and evaluating performance of a Haar feature-based cascade classifier. Similarly, Howell et al. use the NV-IPM image generation tool for designing a camera to optimize for face recognition with the PITT-PATT algorithm\cite{howell2015face}.

In \cite{zelinski_paper}, a motion detection algorithm is used to evaluate image utility. In \cite{zelinski_thesis}, Zelinski notes the inability of the GIQE to accurately predict NIIRS for sparse aperture systems. He cites this as a reason to instead consider image utility through performance of motion detection and spatial target detection algorithms.

Like in \cite{lemaster2017pybsm}, \cite{yuan2014method} uses a Haar feature-based model for comparing image quality from different sensor designs, but Yuan et al. observe overhead imagery instead of faces. The methods used in that work to simulate imagery from different sensors are less robust than pyBSM, NV-IPM, or our method, involving simple addition of noise, blurring, and image contrast reduction.

We build upon these works by using a large, well-curated overhead image dataset transformed with our own sensor model, and state-of-the-art CNNs instead of traditional hand-crafted feature-based models.

%% file: methods.tex
\section{Methods}
\label{sec:methods}
Our approach to modeling algorithm performance with respect to changing image system parameters is straightforward. First, we develop a code which simulates imagery collected by different overhead image systems. The input to this simulator is existing high quality DigitalGlobe imagery. Then, we vary a parameter of this system, such as the optic focal length, and regenerate the entire target dataset with the simulator. Finally, we train and validate an image recognition model on partitions of this dataset, and measure how validation performance changes as a function of the altered system parameter. This section will clarify details about the components of Figure \ref{fig:system}.

A software framework was created to control parameter management, data processing, machine learning, and result logging. This framework was written in Python3, using PyTorch for training and evaluation of CNNs. The framework has the capability to support parameter variation of the image simulation and learning processes. The process of conducting an experimental trial using the framework is explained in Algorithm \ref{algo:trial}. We describe the components of this framework in the following subsections. 

We provide the full code for our image simulator and experimental framework in\footnote{Code repository for this paper: \url{https://github.com/LLNL/sepsense}}.

\begin{algorithm}
\DontPrintSemicolon 
\KwIn{A set of folds $\mathpzc{F} = \{\mathpzc{f}_1, \mathpzc{f}_2, ...,\mathpzc{f}_k\}$, which evenly partition the data. A set of parameters $P$ which contains all static simulator parameters for the trial. A set of parameter values $V = \{v_1, v_2, ..., v_n \}$, which correspond to the independent variable simulator parameter. The number of epochs to train for, $E$}
\KwOut{A set of metric value results $R$, corresponding to trials conducted for all $\mathpzc{f} \in \mathpzc{F}, v \in V$}
\For{$\mathpzc{f}_i$ \textbf{in} $\mathpzc{F}$} {
  $\mathcal{D}_{\text{train}} \gets \mathpzc{f} \in \mathpzc{F}, \mathpzc{f} \neq \mathpzc{f}_i$\;
  $\mathcal{D}_{\text{test}} \gets \mathpzc{f}_i$\;
  \For{$v_j$ \textbf{in} $V$} {
    $m_0 \gets \text{Model}()$\;
    $s \gets \text{Simulator}(P, v_j)$\;
    $\hat{\mathcal{D}}_{\text{train}} \gets \text{transform}(s, \mathcal{D}_{\text{train}})$\;
  	$\hat{\mathcal{D}}_{\text{test}} \gets \text{transform}(s, \mathcal{D}_{\text{test}})$\;
  	\For{$e \gets 0$ \textbf{to} $E$}{
      $m_{e+1} \gets \text{train}(m_e, \hat{\mathcal{D}}_{\text{train}})$\;
      $R_{ije} \gets \text{test}(m_{e+1}, \hat{\mathcal{D}}_{\text{test}})$\;
    }
  }
}
\Return{$R$}\;
\caption{{\sc Trial} conducts experimental trial}
\label{algo:trial}
\end{algorithm}

\subsection{Dataset}
\begin{figure*}[!t]
\centering
\includegraphics[width=\textwidth]{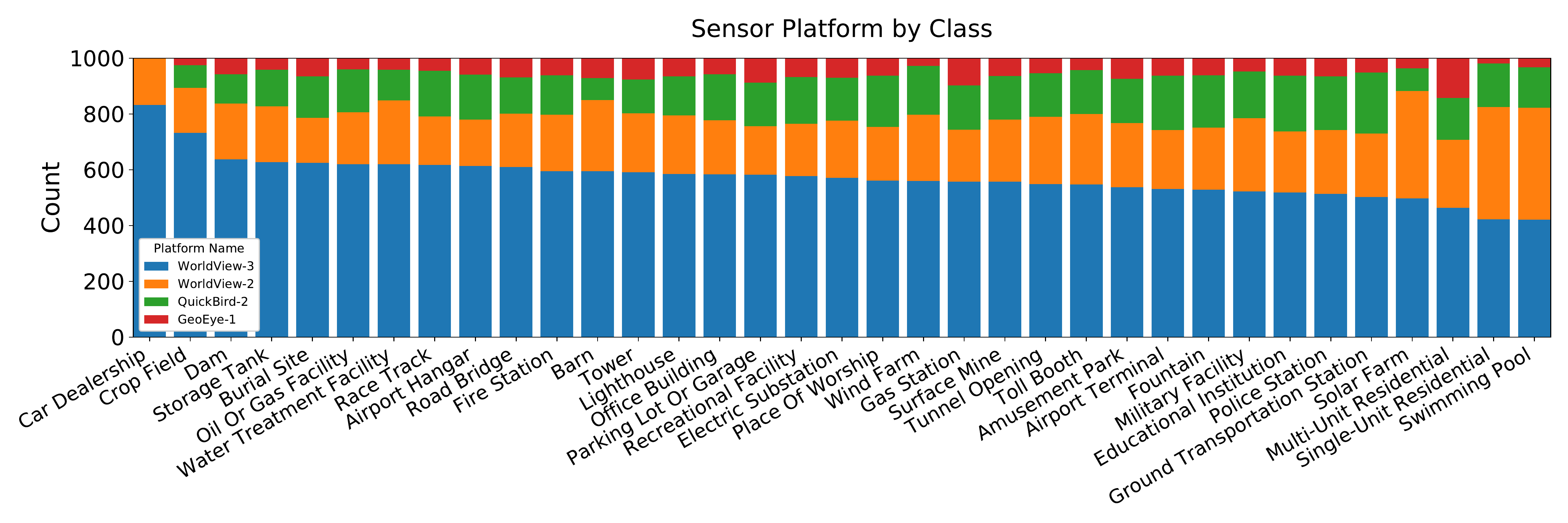} 
\caption{Images from fMoW are collected from four different DigitalGlobe sensor platforms. The proportion of images captured from each platform is relatively even among the 35 classes.}
\label{fig:platform_count}
\end{figure*}

As discussed, the fMoW dataset consists of over one-million images, each of which corresponds to a target from one of 62 classes, or an additional \textit{false detection} class. Images from the dataset were captured in over 200 countries by one of four possible DigitaGlobe sensor platforms: GeoEye1, QuickBird2, WorldView2, or WorldView3. The breakdown of how many images were captured from each platform by class is shown in Figure \ref{fig:platform_count}.

\begin{figure*}[!t]
\centering
\includegraphics[width=\textwidth]{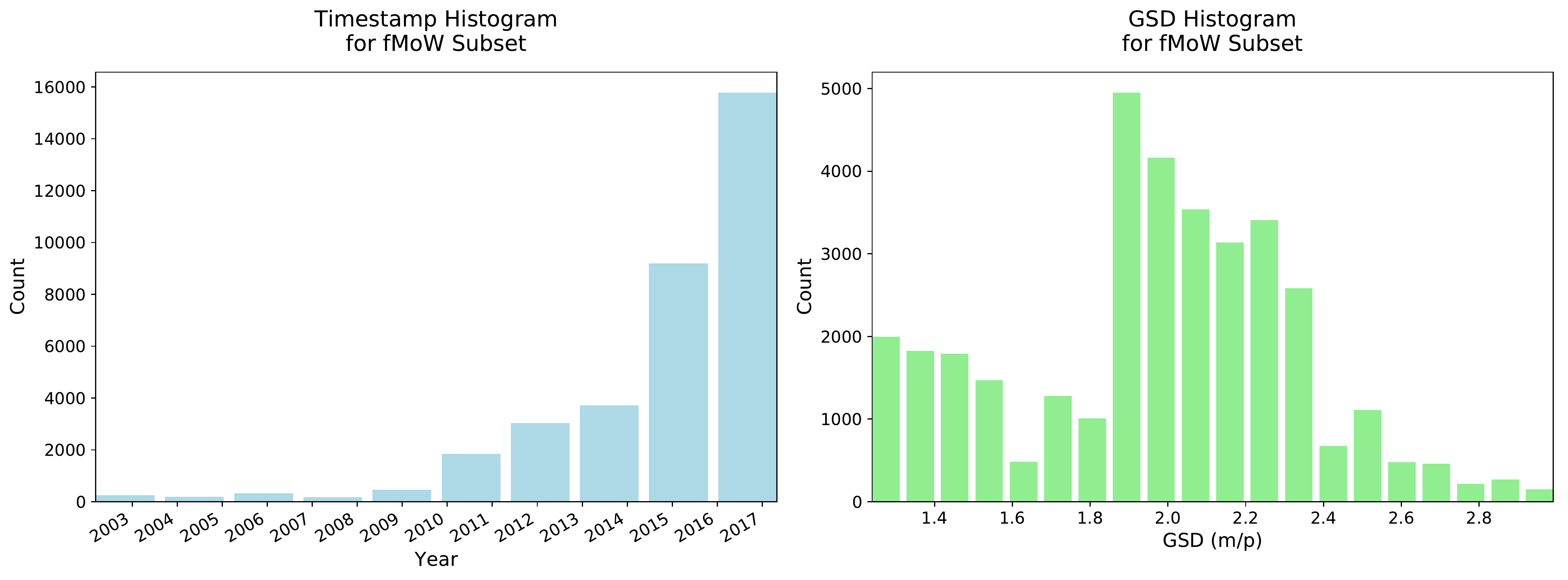} 
\caption{On the left, a histogram is shown for the year during which each image of the fMoW subset was captured. More recent images correspond to WorldView2 and WorldView3, and older images correspond to GeoEye1 and QuickBird2. On the right, a histogram is shown for original GSD values for each image of the fMoW subset. Since all images are regenerated by the same simulated sensing system, and no transformed image can be more resolved than its original version, we are limited by the GSD of the least-resolved image. This also explains the discrepancy in performance between Table \ref{tab:orig_performance} and the experiments using transformed data.}
\label{fig:fmow_hist}
\end{figure*}

\begin{figure*}[!t]
\centering
\includegraphics[width=\textwidth]{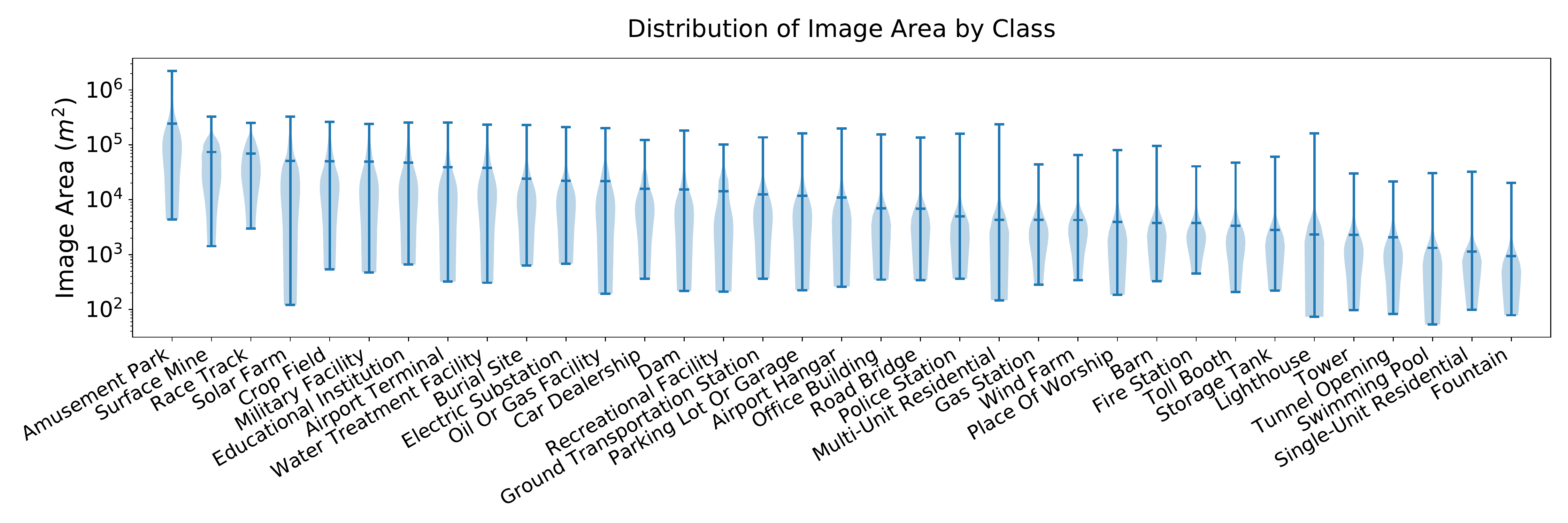} 
\caption{Statistics for the area in meters taken up by images from each class are computed by dividing the number of pixels by the GSD$^2$. Notice the large variation in size, from $10^2$ - $10^6$ $m^2$.}
\label{fig:fmow_area}
\end{figure*}

A unique feature of this dataset is that the target bounding boxes vary widely in area, from {\raise.17ex\hbox{$\scriptstyle\sim$}}10 pixels at a minimum, to {\raise.17ex\hbox{$\scriptstyle\sim$}}700k pixels at a maximum. This presents a unique challenge when considering how to pre-process the data such that a fixed-input model can accept it. We consider two simple approaches to this pre-processing, treated as an experimental variable. Statistics for the dataset are shown in Figures \ref{fig:fmow_hist} and \ref{fig:fmow_area}. Other challenges present in the dataset include artifacts, occlusions, and labeling ambiguities.

The creators of fMoW have provided critical metadata from the original DigitalGlobe imagery, including the \textit{absolute calibration factor}, which is required for converting the data from digital numbers to units of radiance. We use a subset of the total data, comprising 1,000 images from each of 35 of the 62+1 classes. This subset was selected to include only classes with at least 1,000 images which could be transformed for the desired sensor parameter ranges. We use the multispectral (MS) imagery, and not the pan-sharpened RGB imagery, as only the MS imagery can be transformed with our system. Specifically, we use the red, green, and blue MS bands, as this data is most appropriate for standard pre-trained models. In addition, the dataset contains multiple images for each instance of an object, but we use only the first such image for our experiments. We refer to the portion of fMoW used for our study as the \textit{fMoW subset} in the remainder of the text.

\subsection{Data Processing and Calibration}
\label{subsec:dataproc}
The DigitalGlobe satellites undergo an initial calibration to correct for geometric non-uniformity, detector response, lens falloff, and particulate contamination on the focal plane. This \textit{relative radiometric correction} process removes artifacts that produce streaks or banding in the imagery. This is the level of processing at which fMoW data is provided [units of digital numbers]. It should be noted that focal planes aboard DigitalGlobe satellites have adjustable gain and offsets (exposure) settings to accommodate viewing under different illumination and scene reflectance conditions. What this means for users who hope to leverage the radiometric properties of the dataset is that the data must be further calibrated into units of radiance or reflectance.

The 2$^{\text{nd}}$ release of the fMoW dataset included spectral band absolute calibration factor values, which when combined with the other static calibration values, produce an accurate calibration to at-aperture radiance [$W \mu m^{-1} m^{-2} sr^{-1}$] as well as top-of-atmosphere (TOA) reflectance [unitless]. The calibration process and static fine-tuning calibration values for WV-III data can be found in\footnote{\label{link:note1}Radiometric Use of WorldView-3 Imagery:\\\url{https://dg-cms-uploads-production.s3.amazonaws.com/uploads/document/file/207/Radiometric_Use_of_WorldView-3_v2.pdf}}. For the WV-II, Quickbird-II, and Ikonos satellites, these values can be found in\footnote{Radiometric Use of WorldView-2 Imagery:\\\url{http://www.pancroma.com/downloads/Radiometric_Use_of_WorldView-2_Imagery.pdf}},\footnote{Radiance Conversion of QuickBird Data:\\\url{https://apollomapping.com/wp-content/user_uploads/2011/09/Radiance_Conversion_of_QuickBird_Data.pdf}}, and\footnote{IKONOS DN to Radiance Conversion code:\\\url{https://github.com/NikosAlexandris/i.ikonos.toar/blob/master/i.ikonos.toar.py}}.

From units of radiance, the data can be further processed to TOA reflectance, which will compensate for the time of year dependent Earth-Sun distance and solar zenith angle. This conversion is described in\footnotemark[\getrefnumber{link:note1}]. Additional processing which compensates for the atmosphere and surface topology could be beneficial, but it is outside the scope of this effort. 

\subsection{Sensor Modeling}
\label{subsec:sensormodeling}
Once fMoW data is converted to TOA (or at-aperture) radiance, it is possible to use this data to simulate image collection from different remote sensing systems. This section will discuss the approach used here. A more in-depth treatment of the content discussed here is found in \cite{fiete1999image, easton2010fourier, schott2007remote, goodman2008introduction}. 

We begin by noting that one is limited by the capabilities of the image system used to collect the initial dataset. We can only produce imagery of lower quality (lower resolution, higher noise). To summarize this process, the original images undergo a blurring function by applying the optical Modular Transfer Function (MTF). The data is resampled in Fourier space with aliasing artifacts included. The resulting Fourier image is inverse transformed back to image space. A scalar value is then determined that converts the at-aperture radiance images into units of electrons at the detector. The resulting electron image then has photon noise and read noise added. Finally, a sensor gain is applied and the image is quantized to the model's bit resolution. The image can then be converted from digital numbers back to TOA radiance or TOA reflectance from this point. We ignore additional noise sources for simplicity.

\subsubsection{Image Blurring}
We first define the optical cut-off frequency for a diffraction limited incoherent imaging system as: 
\begin{equation}
\nu_{\text{optcut}} = \frac{1}{\lambda \text{FN}} = \frac{D}{\lambda f}
\end{equation}
where $\lambda$ is the wavelength of light, $D$ is the aperture diameter, $f$ is focal length, and FN or 'f-number' is $\frac{f}{D}$. This is the highest frequency sinusoid an optical system can produce on the image plane. This value should be compared directly with the Nyquist sampling limit of the detector. If the frequency is higher than that of the Nyquist sampling limit, it is possible to incur aliasing artifacts, and if it is lower, then the image system will oversample, creating more data than needed and likely at higher noise levels. It should be noted that the lambda value used for these calculations is the shortest wavelength used, which in our work is the center of blue channel. 

The optical Ground Sample Size ($\text{GSS}_{\text{optics}}$) used here follows the Fiete definition of:
\begin{equation}
\text{GSS}_{\text{optics}} = \lambda \frac{H}{D}
\end{equation}
where $H$ is the altitude of the remote sensing system.  
The optical cut-off frequency on the ground is defined as: 
\begin{equation}
\nu_{\text{optcut\_gnd}} = \frac{1}{2 \cdot \text{GSS}} = \frac{D}{2 \cdot \lambda \cdot H}
\end{equation}
The Nyquist sampling limit of our modeled detector is defined as:
\begin{equation}
\nu_{\text{nyquist}} = \frac{1}{2 p}
\end{equation}
where p is the width between adjacent pixel centers on the sensor (pixel pitch). 
And the Nyquist frequency projected to the ground is defined as:
\begin{equation}
\nu_{\text{nyquist\_gnd}} = \frac{1}{2 \cdot \text{GSD}}
\end{equation}
The maximum recorded frequency $\nu_{\text{cutoff}}$ will depend on whether the system over or under samples the image plane. So: 
if $\nu_{\text{optcut}} > \nu_{\text{nyquist}}$, then $\nu_{\text{cutoff\_gnd}} = \nu_{\text{optcut\_gnd}}$, and $\nu_{\text{cutoff}} = \nu_{\text{optcut}}$, else $\nu_{\text{cutoff\_gnd}} = \nu_{\text{nyquist\_gnd}}$ and $\nu_{\text{cutoff}} = \nu_{\text{nyquist}}$. 

To connect the optical and sampling parameters of the modeled telescope with the DigitalGlobe image, we first define the DigitalGlobe imagery nyquist sampling limit on the ground as: 
\begin{equation}
\nu_{\text{DG\_nyquist\_gnd}} = \frac{1}{2 \cdot \text{GSD}_{\text{DG}}}
\end{equation}
The DigitalGlobe frequency array is therefore defined as:
\begin{equation}
\xi_{\text{DG}} =
\begin{bmatrix}
-\nu, &
-\nu + 2 \cdot \frac{\nu}{N}, &
\hdots, &
0, &
\hdots, &
\nu
\end{bmatrix}
\end{equation}
where $\nu$ is $\nu_{\text{DG\_nyquist\_gnd}}$ and $N$ is the number of pixels in the original image.

Finding the indices of the closest value to $\nu_{\text{cutoff\_gnd}}$ in $|\xi_{\text{DG}}|$ will allow us to resample the imagery in Fourier space to the dimensions of the image plane defined in units relative to the $\nu_{\text{cutoff}}$. This is done by cropping the image in Fourier space at these indices, thus discarding spatial frequency data above $|\nu_{\text{cutoff\_gnd}}|$. A full description of the cropping, padding, and frequency wrapping that is necessary to accurately model the range of artifacts caused by over and under sampling is beyond the scope of this document. This was carried out and tested in this work.

Blurring the imagery is done by applying the system MTF. The system MTF is defined as the normalized autocorrelation of the scaled pupil function. Here we define the pupil function $p(x, y)$ as a complex-valued circle function with 1 in the real-part and 0 in the imaginary part. And $P \left[ \frac{x}{\lambda f}, \frac{y}{\lambda f} \right]$ is the Fourier Transform of the pupil function. 
The optical transfer function (OTF) \cite{easton2010fourier} is defined as:
\begin{equation}
\mathcal{OTF}[\xi, \eta] = \mathcal{F} \left\lbrace \left| P \left( \frac{x}{\lambda f}, \frac{y}{\lambda f} \right) \right|^2 \right\rbrace
\end{equation}
where $\mathcal{F}$ denotes the Fourier transform.
The MTF is the OTF normalized by its central ordinate:
\begin{equation}
\mathcal{MTF}[\xi, \eta] \equiv \frac{\mathcal{OTF}[\xi, \eta]}{\mathcal{OTF}[\xi, \eta]_{\xi=0, \eta=0}}
\end{equation}
Using the Fourier Filter Theorem the MTF can be used to degrade the image in Fourier space via a simple multiplication:
\begin{equation}
I[\xi, \eta] = \mathcal{MTF}[\xi, \eta]I_0[\xi, \eta]
\end{equation}
where $I[\xi, \eta]$ is the spatially degraded image and $I_0[\xi, \eta]$ is the original image, both of which are in Fourier space. The resulting image is then resampled by cropping in Fourier space
and then (if needed) summing aliased high frequencies onto their corresponding low frequency counterparts. The resulting array is then inverse-transformed back to the spatial domain, $I(x,y)$. 

\subsubsection{Radiometry}
$I(x,y)$ is in units of at-aperture radiance at this point in our treatment. This section will discuss how to take the spatially-degraded image and degrade it further by shot noise and read noise.

Before the shot noise can be determined the image must be converted to units of electrons on the sensor. This is done by first calculating the spectral flux entering each pixel. A flux scalar is given by:
\begin{equation}
\alpha = \Delta \lambda \cdot \tau \cdot A_{\text{pixel}} \cdot \frac{\pi}{ 1 + 4 \cdot \text{FN}^{2}}
\end{equation}
where $\Delta \lambda$ is the spectral band width, $\tau$ is the optical transmission fraction, and $A_{\text{pixel}}$ is the area of the pixel. An electron scalar, denoted by $\beta$, can be calculated by using the detector’s quantum efficiency, integration time, and energy per photon. This number can then be converted to electrons by dividing by the amount of energy contained within a photon:
\begin{equation}
\beta = \text{QE} \cdot T_{\text{int}} \cdot \alpha \cdot \left( \frac{h c}{\lambda} \right)^{-1}
\end{equation}
where $h$ is Planck’s constant ($h = 6.6260 \cdot 10^{-34} [J/s]$) and $c$ is the speed of light ($c = 2.9979 \cdot 10^8 [m/s]$). The image expressed in units of electrons is: 
\begin{equation}
I_{e}(x,y) = I(x,y) \cdot \beta
\end{equation}

Shot noise values can be created for each xy-pixel by sampling from a Poisson distribution with a lambda value of $\sqrt{I_{e}(x,y)}$. Because lambda is generally relatively large, a Gaussian function can be used for the random distribution. Photon noise can be computed as:
\begin{equation}
N_{\text{photon}} = \mathcal{N}(\mu=0, \sigma^2=1) \sqrt{I_{e}(x,y)}
\end{equation}

The read noise is defined by the focal plane array manufacturer in units of electrons. It is Gaussian in nature. The total noise per pixel is the summation of these values:
\small
\begin{equation}
N(x,y) =  \mathcal{N}(\mu=0, \sigma^2=1) \sqrt{I_{e}(x,y)} + \mathcal{N}(\mu=0, \sigma^2=1) \sigma_{\text{read}}
\end{equation}
\normalsize

The output image from the model is therefore expressed in volts as:
\begin{equation}
I_{\text{volts}}(x,y) = \frac{1}{G} \cdot (I_{e}(x,y)  + N(x,y))
\end{equation}
where the gain, $G$, is defined as $\frac{\epsilon}{2^{n}}$, $\epsilon$ is the \textit{electron well depth}, and $n$ is the \textit{bit depth}. The resulting image can then be quantized.

These simulated images can be converted back to radiance by multiplying by $\frac{G}{\alpha\beta}$. TOA reflectance can be calculated from this point.

\subsubsection{Example Images}
\begin{figure*}[!t]
\centering
\includegraphics[width=\textwidth,draft=false]{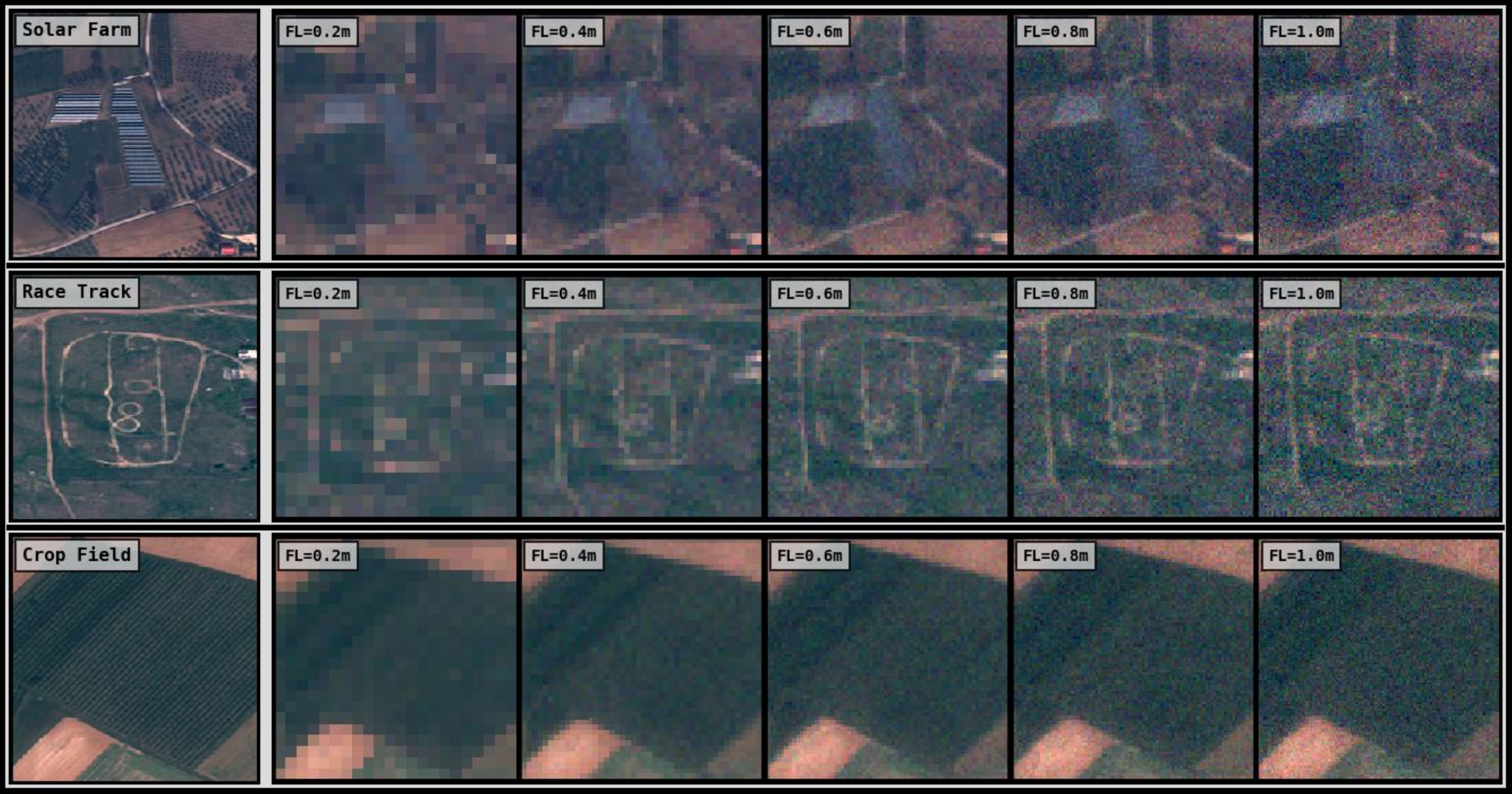} 
\caption{Three example image classes from the fMoW subset are transformed with varying focal length, with other parameters kept constant. Original images are on the left separated by a gray column, with their class labels shown in the upper left. }
\label{fig:fmow_viz}
\end{figure*}

\begin{figure*}[!t]
\centering
\includegraphics[width=\textwidth,draft=false]{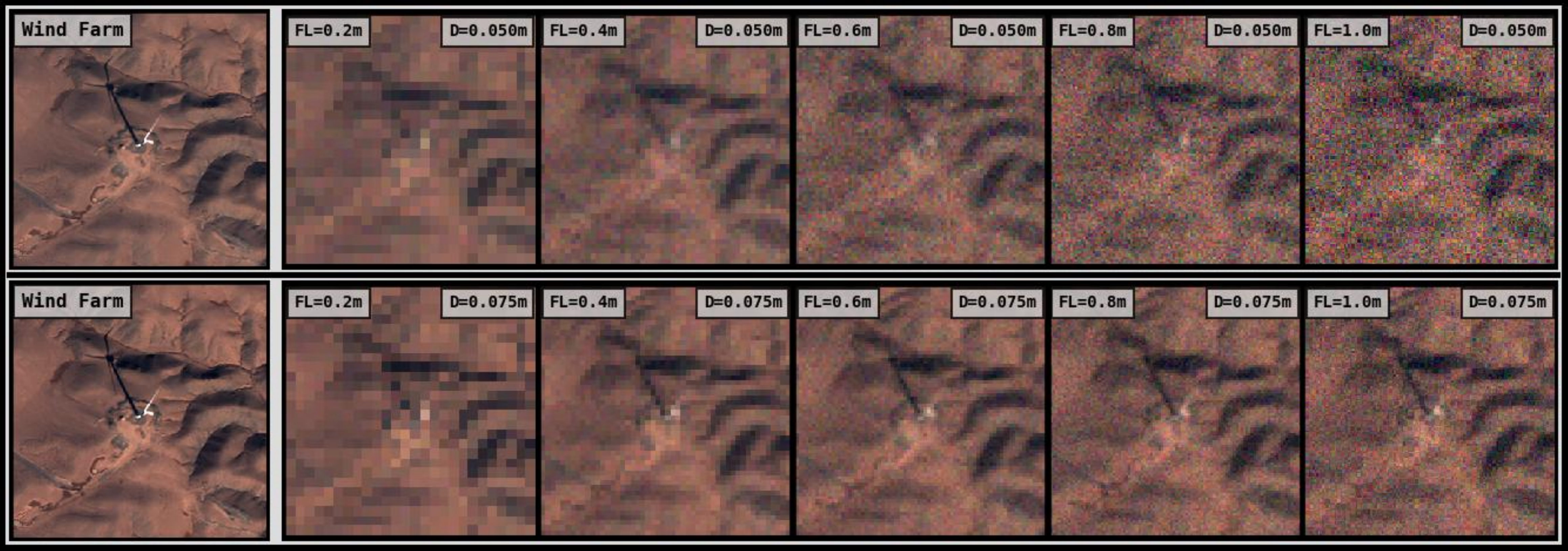} 
\caption{An example wind turbine image from the fMoW subset is transformed with increasing focal length (left to right) and increasing aperture diameter (top to bottom). The original image is shown on the left (repeated), separated by a gray column.}
\label{fig:d_compare}
\end{figure*}

The effects of this transformation process are exhibited for images from the fMoW subset in Figures \ref{fig:fmow_viz} and \ref{fig:d_compare}. All transformations were done using parameter combinations from the experiments described in Section \ref{sec:results}.

In Figure \ref{fig:fmow_viz}, three example images are transformed with varying focal length. Since the simulator is only capable of producing images with less information (i.e. blurring, noise) than the originals, all transformed images should appear degraded with respect to their original counterparts. For instance, notice that fine-grain details, such as striations between cells in the solar farm, or small roads in the race track, are lost in all transformed images. On the other hand, notice how little interpretability of certain classes may be affected, such as the crop field in the third row. With respect to changing focal length, notice that the GSD decreases while noise artifacts increase.  

In Figure \ref{fig:d_compare}, a single example image is transformed with varying focal length and varying aperture diameter. Notice that the image transformed with $f=1.0m, D=0.75m$ has significantly less noise than the image transformed with $f=1.0m, D=0.50m$. This is because more light is admitted to the system with larger aperture diameter, meaning the light hitting a given CCD cell is less likely to fall below the read noise floor. In contrast, there is little difference between the two images with $f=0.2m$.

Note that all example images have been resampled to a fixed size using nearest-neighbors interpolation. While the images would appear more natural if resampled with a bilinear or bicubic filter, the nearest-neighbors visualization gives a nice representation of how the number of ``pixels on target'' varies with focal length. For the actual experiments, bilinear interpolation was used.

\subsection{CNN Architectures}
\label{subsec:cnn}
We performed experiments with four CNN architectures: SqueezeNet \cite{iandola2016squeezenet}, VGG \cite{Simonyan15}, ResNet \cite{he2016deep}, and DenseNet \cite{huang2017densely}. These architectures were selected to exhibit the state-of-the-art in CNNs for visual recognition, while capturing significant variation in design and parameter count. All models used are implemented in the PyTorch torchvision library\footnote{PyTorch torchvision documentation:\\\url{https://pytorch.org/docs/stable/torchvision/models.html}}, and are trained using either cross-entropy loss or triplet margin loss. Minor modifications were made to the original torchvision models to correspond to the different performance metrics and learning objectives, described in Section \ref{subsec:mod_mod}.

The VGG model was one of the first successful CNN architectures, and is known for its simple structure and high parameter count. The strategy of VGG is to use many repeated 3x3 convolutional layers, with 2x2 max pooling layers to reduce the spatial dimension between blocks. While it has been outclassed in performance by ResNet, DenseNet, and other architectures, it remains useful as a reliable baseline. We use the popular 16-layer variant of this model (VGG16). 

A key issue with VGG is that the vanishing gradient problem manifests as more layers are added. The ResNet architecture addresses this problem by introducing the \textit{skip connection} structure: a basic block which has its input added to its output. This allows for successful optimization with a much deeper network. We experiment with the 152-layer version of the ResNet architecture (ResNet152). 

The DenseNet architecture takes the skip connection concept of ResNet further by concatenating layer inputs to layer outputs, instead of adding them. This model encourages feature reuse by directly connecting all layers of the network in this fashion. We utilize a 161-layer DenseNet (DensetNet161) architecture, which has performed well on ImageNet and on the fMoW dataset. The creators of fMoW use DenseNet161 as their baseline architecture \cite{fmow}, and all of the top placing solutions in the fMoW challenge utilized some variant of DenseNet in their models\footnote{fMoW GitHub repository: \url{https://github.com/fMoW}}. The architecture is memory-intensive, but trains in very few epochs in practice.

In contrast to the other models, the SqueezeNet architecture is designed to have few parameters and small size. We use an optimized variant of this architecture, dubbed SqueezeNet1.1\footnote{Official SqueezeNet repository:\\\url{https://github.com/DeepScale/SqueezeNet/tree/ master/SqueezeNet\_v1.1}}. SqueezeNet achieves reasonable performance on ImageNet with more than 20x fewer parameters than DenseNet161, shown in Table \ref{tab:arch_info}. Architectures with small size like SqueezeNet may be particularly interesting for remote sensing scenarios, in which a model could be deployed onboard a sensing platform. In this case, small memory footprint and low power consumption constraints are critical.

\subsubsection{Transfer Learning}
In the context of training CNN models for visual recognition, transfer learning usually refers to the process of training a model on a large standard image dataset, then fine-tuning the model on the image data of interest, possibly from a different domain. In practice, transfer learning is always used for solving natural imagery visual recognition problems. In particular, the ImageNet dataset is often used to pre-train the target model. For our experiments, all CNN models were pre-trained on ImageNet unless otherwise specified. These pre-trained weights were downloaded through the torchvision library.

\begin{table}
\centering
\captionof{table}{CNN Architecture Information} \label{tab:arch_info} 
\begin{tabular}{lrrr}
\toprule[1.5pt]
\textbf{Architecture} & \textbf{Num Params}* & \textbf{Top-1 Accuracy} & \textbf{Top-5 Accuracy} \\
\midrule
SqueezeNet1.1 & 1.2M & 0.582 & 0.806 \\
VGG16 & 138.4M & 0.716 & 0.904 \\
ResNet152 &  60.2M & 0.783 & 0.941 \\
DenseNet161 &  28.7M & 0.777 & 0.938 \\
\bottomrule
\end{tabular}
\begin{tablenotes}
\item[*]*Parameter counts were gathered directly from the actual torchvision models used. Model statistics shown are ImageNet 1-crop error rates (224x224). Stats from \url{https://pytorch.org/docs/stable/torchvision/models.html}.
\end{tablenotes}
\end{table}

\subsection{Performance Evaluation}
We consider two image recognition tasks to evaluate machine image interpretability as a function of varying sensor parameters. The first of these tasks is the classification problem, which requires selecting the correct class from a known set of classes for a given instance. The second of these tasks is the retrieval problem, which requires selecting instances matching a \textit{probe} class from a set of instances comprising a \textit{gallery}.

We use classification and retrieval specifically for three reasons. First, they are commonly used to solve real-world problems. Hence, results from these experiments can be directly used to assess real recognition software in use on overhead imagery. Second, these problems are relatively simple, unlike the detection and segmentation problems, which involve more moving parts. We want to identify a simple relationship, which can then be used as a building block to understand more complex recognition problems. Finally, the retrieval problem in particular allows us to develop a generalizable method, which can be applied to unseen classes. In short, a \textit{salient feature extractor} can be trained on a known dataset of images, and later tested on new data with unknown classes, without retraining.

\subsubsection{Classification}
We evaluate four metrics for the classification problem: top-1 accuracy, top-3 accuracy, Area Under the (ROC) Curve (AUC), and Average Precision (AP). 

Top-1 and top-k accuracy are the most basic metrics available for evaluating classification, but are also easily interpretable. For this reason, they may be of interest for the sensor design process. We chose top-3 accuracy by analyzing the datset, and determining that certain groups of up to three objects were easily confused or arbitrarily labeled. The implication is that we are still interested in a model which can predict three candidate classes, one of which will be correct for the input image.

Use of Average Precision (AP) is common for machine learning applications, because it supplies a principled summary metric of precision at all levels of recall:
\begin{equation}
\label{eq:ap}
\text{AP} = \int_0^1 p(r)dr
\end{equation}
where $r$ is recall, and $p(r)$ is precision at the given recall value. In practice, a given application may attach more or less importance to precision or recall, but AP gives a nice summary in the case where no prior has yet been determined.

Alternatively, the Area Under the (Receiver Operator Characteristic) Curve (AUC) may be beneficial, if the application benefits from a summary of True Positive Rate (TPR) a.k.a. recall vs. False Positive Rate (FPR). The Precision-Recall curve and AP are commonly used when the target dataset has a class imbalance, whereas the ROC curve and AUC are more appropriate when each class has a similar number of samples. This is because the ROC curve may inflate the results in the case of class imbalance\cite{davis_roc}.

\subsubsection{Retrieval}
We also utilize the AP metric to quantify retrieval performance, because it can be evaluated using any arbitrary feature space, so long as some distance metric can be measured between elements in the space. For the retrieval problem, we consider a different interpretation of AP\cite{kishida2005property} than the one introduced in Equation \ref{eq:ap}. To measure the AP of a retrieval query, we first measure distances between the probe image and all gallery images. Then, the gallery elements are ranked from least to greatest distance from the probe. A gallery image \textit{matches} a probe image if they share the same object category or \textit{class}. To measure the precision for the top-ranked $m$ images, we take:
\begin{equation}
\label{eq:precision_retrieval}
p_m = \frac{1}{m}\sum_{k=1}^{m}x_k
\end{equation}
where
\begin{equation}
x_k =
\begin{cases}
    1			& \text{if k}^{\text{th}} \text{ gallery image matches probe} \\
    0           & \text{otherwise}
\end{cases}
\end{equation}
If $R$ represents the total number of gallery images, then:
\begin{equation}
\label{eq:ap_retrieval}
\text{AP} = \frac{1}{R}\sum_{i=1}^{n}x_i p_i = \frac{1}{R}\sum_{i=1}^{n}\frac{x_i}{i}\sum_{k=1}^{i}x_k
\end{equation}
Note that we use 2-norm distance in a CNN feature space to measure distance between images.

\begin{figure*}[!t]
\centering
\includegraphics[width=\textwidth,draft=false]{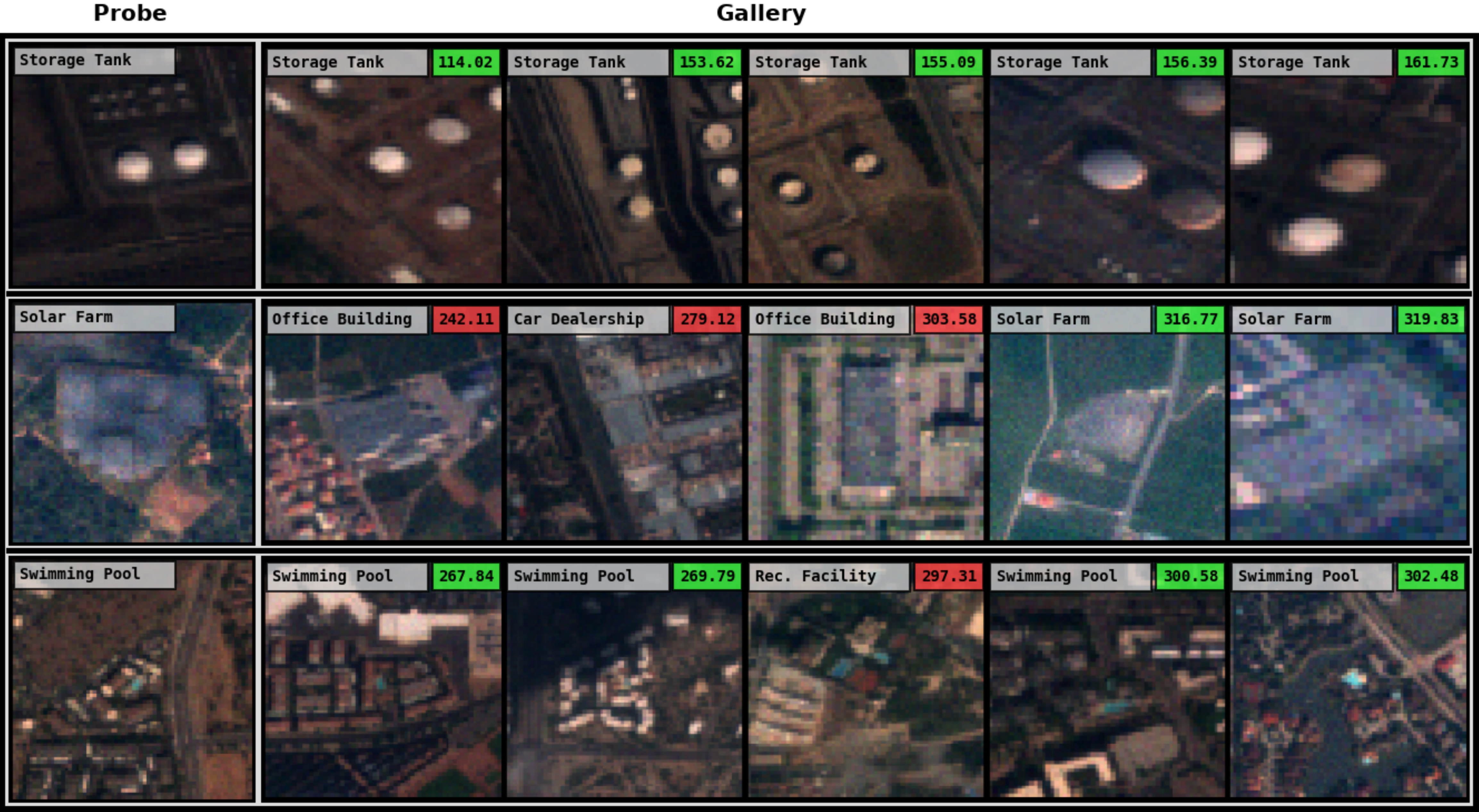} 
\caption{Retrieval examples using the baseline model with a single fold of the fMoW subset. Each row of images represents a single retrieval query, with the true class of each image shown in its upper left. Probe images are indicated on the left of each row, separated from the gallery by a gray bar. Gallery images are ranked by least to greatest distance from the probe, left to right. The 2-norm distance from the probe is shown in the upper right of each gallery image, colored green if the image class matches the probe class, and red if it does not.}
\label{fig:retrieval}
\end{figure*}

AP allows us to measure the utility of pre-trained model features, and the utility of fine-tuned features. Using the retrieval problem in particular, we can evaluate the \textit{zero-shot} scenario, in which no samples of the target classes have been seen during training. In Figure \ref{fig:retrieval}, the retrieval problem is visualized for an experiment conducted on the fMoW subset. Notice that top-ranked gallery samples share common features with the probe, even when their labels do not match.

In the remaining text, AP will be denoted as cAP when referring specifically to the classification problem, and rAP when referring to the retrieval problem. In addition, note that our AP values are computed as an average across all possible probe-gallery combinations, with all 35 classes, unless otherwise indicated. This is typically denoted as mAP, but we will not use this notation.

\subsection{Learning Objective}
\label{subsec:learning_objective}
We consider two different learning objectives to train the CNN models: \textit{cross-entropy loss} and \textit{triplet margin loss}. 

\subsubsection{Cross-Entropy Loss}
Cross-entropy loss is considered the standard objective function for training a neural network to solve the classification problem. This problem and objective are typically effective when there is a known set of classes with many samples, and all future test samples belong to one of these known classes. 

\subsubsection{Triplet Margin Loss}
In contrast, triplet margin loss is a more popular objective for the retrieval problem. Triplet margin loss is an example of a \textit{metric learning} objective. In metric learning, the goal is to find a semantically-meaningful feature space based on some target distance metric. Elements projected into the feature space by some function, in this case a CNN model, may be called \textit{features} or \textit{embeddings}. Image embeddings from the same class should be closer together in the learned feature space, whereas embeddings from different classes should be further apart. The triplet margin loss operates on sets of three embeddings: an \textit{anchor} and \textit{positive} embedding from the same class, and a \textit{negative} embedding from a different class. The loss function is defined as follows, with $a$, $p$, and $n$ corresponding to the \textit{anchor}, \textit{positive}, and \textit{negative} embeddings respectively:
\begin{equation}
L(a, p, n) = \max \{d(a_i, p_i) - d(a_i, n_i) + {\rm margin}, 0\}
\end{equation}
where
\begin{equation}
d(x_i, y_i) = \left\lVert {\bf x}_i - {\bf y}_i \right\rVert_p
\end{equation}
We follow the triplet learning method described by Schroff et al. in \cite{schroff2015facenet}. In this paper, 2-norm distance is used for the distance metric, and each embedding is (2-norm) normalized to magnitude 1 prior to computation of the obective. A value of 0.2 is used for the margin, which we have also found is a reliable default.

Triplets are sampled such that 20 images from each of 4 classes are seen during each batch. When each image has been sampled in a triplet exactly once, one epoch has been completed. We note that training is not as stable using this objective, and typically requires more epochs to optimize than cross-entropy loss. 

\subsection{Model Modifications}
\label{subsec:mod_mod}
To train with cross-entropy loss, the original fully-connected classification layer in each CNN model was replaced by a randomly-initialized fully-connected layer with outputs corresponding to the number of classes used (35 for most experiments).

To compute rAP or train with triplet margin loss, image embeddings must be extracted from the target CNN models prior to the classification layer. In our models, this was done either by extracting the embeddings directly from the second to last layer, or by adding an additional 128-output fully-connected layer prior to the last layer, and extracting from that. The DenseNet161 and ResNet152 models were left unchanged for cross-entropy loss training, while VGG16 and SqueezeNet1.1 added the additional 128-output layer. This is because the features at the end of the original VGG16 and SqueezeNet1.1 models are too high-dimensional and therefore less effective for the retrieval problem.

For triplet-margin loss training, we also added the additional 128-output layer to the end of the DenseNet161 model. It is import to use a lower-dimensional space for the learned embedding space to avoid the \textit{curse of dimensionality} \cite{kulis2013metric}. 

\subsection{Pre-Processing}
\label{subsec:preproc}
A significant set of operations was applied to the image data both before and after the simulation code was applied.

\subsubsection{Before Simulation}
First, the R, G, and B bands were extracted from the DigitalGlobe 16-bit MS images comprising the fMoW subset. Then, the images were either cropped or resized to $224 \times 224$ pixels, such that the centroid of the annotated object in the image was at the center of the processed output image. Note that no resizing was applied in the case of the cropping operation, meaning that objects larger than $224 \times 224$ pixels would lose any of those additional pixels. Further, objects smaller than $224 \times 224$ pixels would also contain whatever additional pixels were outside the object bounding box and still within the $224 \times 224$ box. In the case where one or both dimensions of the full image was smaller than $224 \times 224$ pixels, the remaining space was zero-padded. In the case of the resize operation, the object bounding box would be resized to $224 \times 224$ pixels without preserving aspect ratio, using bilinear interpolation. Note that the cropping operation could benefit smaller objects, since additional context would be included, while the resizing operation could benefit larger objects, since the entire spatial extent would be within the final image (despite many original pixels being lost). Both operations are flawed for many of the images/objects, but we will show that the cropping operation is better on average.

An additional pre-processing step was required for the resizing operation, involving padding the edges with ``reflected'' values. This was done to avoid boundary artifacts which occurred after converting to and from the frequency domain with DFT.

After the images were cropped or resized, they were translated into units of at-aperture radiance and put into the sensor simulation code. The output imagery produced was in units of TOA reflectance.

\subsubsection{After Simulation}
\label{subsec:aftersim}
The imagery in units of TOA reflectance could be of different dimension than the input $224 \times 224$ pixels, so resampling to $224 \times 224$ pixels was necessary, for which bilinear interpolation was again used. In the case of resized images, the additional padding previously mentioned was clipped off after this interpolation. Since the domain of TOA reflectance is close to [0, 1], additional normalization of the domain is not necessary. Initial experiments were conducted with \textit{standard scores} (mean subtraction and standard deviation division), but this was found to be beneficial only for the scenario in which a pre-trained model was not fine-tuned. All experiments with fine-tuning on the fMoW subset have no additional data normalization.

\subsection{Hardware and OS}
All experiments were conducted on machines with NVIDIA GPUs, either Quadro M6000, Pascal Titan X, or Tesla K80. The experiments were conducted inside Docker containers based on Ubuntu 18.04 and CUDA 10.0. Most experiments took around 48 hours to complete with the use of 2 GPUs, and a mini-batch size of 88.

\subsection{Parameters}
All experiments were conducted with a set of parameters which remain constant, as certain other parameters were varied. The set of parameters corresponding to the sensor simulator are found in Table \ref{tab:sensor_parameters}, and the set of parameters corresponding to all other aspects of the experimental design are found in Table \ref{tab:experiment_parameters}.

\bigskip
\begin{table}
\centering
\captionof{table}{Sensor Parameters} \label{tab:sensor_parameters} 
\begin{tabular*}{0.48\textwidth}{l @{\extracolsep{\fill}} rr}
\toprule[1.5pt]
\textbf{Constant Parameters} & \\
\midrule[1.5pt]
\textit{Parameter} & \textit{Value} & \textit{Units}\\
\midrule
Optical Transmission &  [B:0.95, G:0.95, R:0.95] & unitless \\
Altitude & 500e3 & meters \\
Pixel Pitch & 6e-6 & meters \\
Integration Time & 2.5e-4 & seconds \\
Wavelength Freq. & [B:4.5e-7, G:5.5e-7, R:6.5e-7] & meters \\
Spectral Bandwidth & [B:0.1, G:0.1, R:0.1] & meters \\
Electron Well Depth & 40300 & unitless \\
Bit Depth & 13 & unitless \\
Read Noise & 12.5 & unitless \\
Quantum Efficiency & [B:0.22, G:0.22, R:0.16] & unitless \\
\toprule[1.5pt]
\textbf{Variable Parameters} & \\
\midrule[1.5pt]
\textit{Parameter} & \textit{Range} & \textit{Units}\\
\midrule
Focal Length & 0.1 - 1.0 & meters \\
Aperture Diameter & 0.050 - 0.075 & meters \\
\bottomrule
\end{tabular*}
\end{table}

\bigskip
\begin{table}
\centering
\captionof{table}{Experimental Parameters} \label{tab:experiment_parameters} 
\begin{tabular*}{0.48\textwidth}{l @{\extracolsep{\fill}} r}
\midrule[1.5pt]
\textbf{Constant Parameters} & \\
\toprule[1.5pt]
\textit{Parameter} & \textit{Value} \\
\midrule
Image Size & 224x224x3 \\
Optimizer &  Adam \\
Learning Rate &  1e-4 \\
Cross Validation & 1x10 Folds \\
Batch Size* & \{80, 88\} \\
Num GPUs* & 0-2 \\
\midrule[1.5pt]
\textbf{Variable Parameters} & \\
\toprule[1.5pt]
\textit{Parameter} & \textit{Range} \\
\midrule
Pretrained &  \{True, False\} \\
CNN Architecture &  \{SqueezeNet1.1, VGG16,\\ & ResNet152, DenseNet161\} \\
Learning Objective & \{Cross-Entropy, Triplet Margin\} \\
Num Classes &  \{2, 3, 5, 10, 15, 35\} \\
Num Epochs & 0-20 \\
Pre-Processing Method & \{Crop, Resize\} \\
\bottomrule
\end{tabular*}
\begin{tablenotes}
\item[*]*The batch size and number of GPUs were selected based on memory requirements. These parameters were not found to affect the optimization within the reported range.
\end{tablenotes}
\end{table}

\subsection{CNN Experiments}
\label{subsec:experiments}
Three main types of CNN experiments were conducted using the described methods. All experiments were conducted using 10-fold cross-validation with 35 classes, training with 900 images from each class, and validating with 100 images from each class. Within each class, the images were assigned to each fold randomly.

\subsubsection{Basic Training and Evaluation}
The three CNN models were trained and validated on the original dataset, establishing an upper bound on performance.

\subsubsection{Degraded Training and Evaluation}
For each value of some sensor parameter, the entire dataset was transformed, and the CNN models were trained and validated on the transformed dataset.

\subsubsection{Degraded Training with Zero-Shot Evaluation}
The class space was first divided into two groups of 20 and 15 mutually exclusive classes. Then, the degraded training and evaluation experiment was conducted, training on the group of 20, and validating on the group of 15. For this experiment, only the retrieval problem could be used for evaluation, since the network would be unable to perform classification with the unknown classes.

\subsection{IQA Experiments}
\label{subsec:iqa_experiments}
Both full-reference and no-reference IQA metrics were measured for comparison against CNN performance and GIQE5 NIIRS. Each metric was computed for all 1,000 transformed images across each of 35 classes, for each simulator parameter configuration. Images were pre-processed using the previously described ``cropping'' method, and normalized to the range $[0, 1]$.

\subsubsection{Full-Reference}
The PSNR and SSIM metrics were measured comparitively using a given image \textit{Before Simulation} and \textit{After Simulation} as described in the Pre-Processing section. For SSIM, the score was computed separately for each channel and averaged to produce the final result. The scikit-image\footnote{scikit-image: \url{https://scikit-image.org/}} implementation of these metrics was used for the experiments.

\subsubsection{No-Reference}
The BRISQUE and NIQE metrics were measured using only the \textit{After Simulation} images. For both metrics, the score was computed separately for each channel and averaged to produce the final result, like for SSIM. The scikit-video\footnote{scikit-video: \url{http://www.scikit-video.org/stable/}} implementation of NIQE was used, while the BRISQUE implementation is from\footnote{BRISQUE Implementation:\\\url{https://github.com/spmallick/learnopencv/tree/master/ImageMetrics}}. The performance of these algorithms would likely be improved by retraining their respective models with in-domain data, but this is left to future work.

%% file: results.tex
\section{Results and Analysis}
\label{sec:results}

\subsection{Calibration}
We begin by training the four target CNN architectures on the original fMoW subset, in order to establish an upper bound for performance. We know this will constitute an upper bound, because the transformations applied to the data are a strict degradation involving downsampling and adding noise. We show results on the validation set in Table \ref{tab:orig_performance}. The peak values for five metrics are listed, in addition to the training epoch at which they were achieved. All four models were pre-trained on ImageNet and were fine-tuned for five epochs. Interestingly, the DenseNet161 architecture performs better than ResNet152, despite ResNet152 performing marginally better on the original ImageNet dataset. This showcases again how different architectures may perform better on different datasets. Therefore, it is important to understand the performance relationship here studied for multiple state-of-the-art architectures, as any could be optimal for a given dataset.

\setlength{\heavyrulewidth}{1.5pt}
\setlength{\abovetopsep}{4pt}
\begin{table*}[!htbp]
\centering
\caption{Peak Results for fMoW Subset}
\label{tab:orig_performance}
\begin{tabular}{ccccccccccc}
\toprule
Architecture &  \multicolumn{2}{c}{Top-1 Accuracy} &  \multicolumn{2}{c}{Top-3 Accuracy} &  \multicolumn{2}{c}{AUC} &  \multicolumn{2}{c}{cAP} &  \multicolumn{2}{c}{rAP}\\
\midrule
{} & Epoch & Value & Epoch & Value & Epoch & Value & Epoch & Value & Epoch & Value\\
SqueezeNet1.1 & 13 & 0.476 & 13 & 0.692 & 13 & 0.901 & 13 & 0.391 & 13 & 0.151\\
VGG16 & 3 & 0.579 & 3 & 0.771 & 3 & 0.926 & 5 & 0.518 & 3 & 0.312\\
ResNet152 & 3 & 0.602 & 3 & 0.792 & 3 & 0.947 & 5 & 0.621 & 5 & 0.376\\
DenseNet161 & 3 & 0.623 & 2 & 0.809 & 3 & 0.952 & 4 & 0.650 & 4 & 0.405\\
\bottomrule
\end{tabular}
\end{table*}

\subsection{Baseline}
\begin{figure}[!t]
\centering
\includegraphics[trim={0 0 3cm 0},width=0.5\textwidth,draft=false]{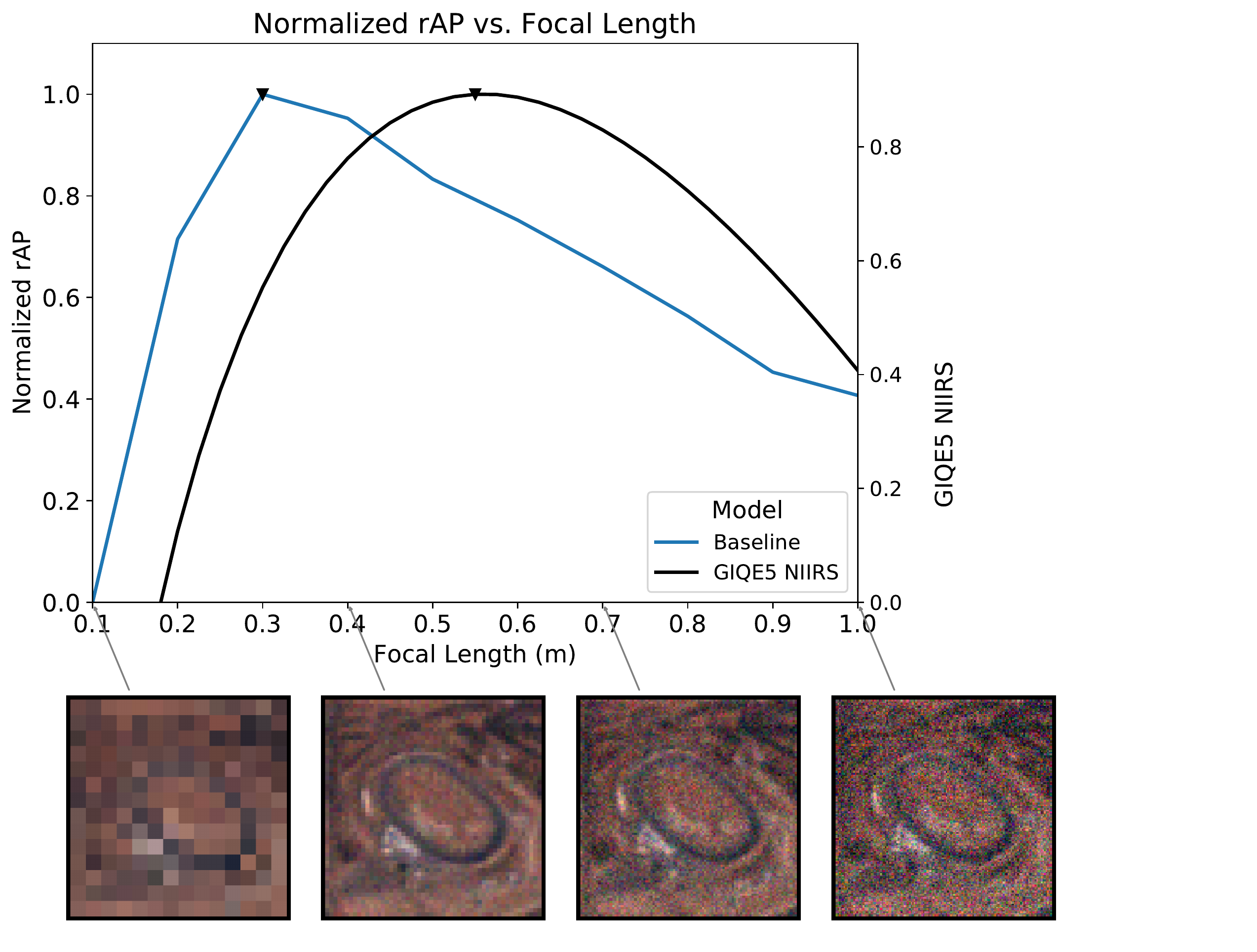} 
\caption{This plot visualizes the baseline experiment of this work, in which rAP is averaged across validation partitions of the fMoW subset, for a range of sensor system focal length values ($D$=0.05m). An example image of a race track is shown for different focal length values to demonstrate the visual changes which result from the transformation. A clear peak is present for rAP, at which point the optimal tradeoff between resolution and noise for the CNN has been reached. The GIQE5 NIIRS curve for these focal length values has a different shape, peaking at a larger focal length value. This shows that CNN performance and human interpretability may not have the same relationship with sensor system parameters. Note that rAP has been normalized to [0, 1], while GIQE5 NIIRS is unaltered, to better compare the peaks between the two curves.}
\label{fig:cornerstone}
\end{figure}

For our baseline experiment, we consider the relationship between rAP of DenseNet161 and optic focal length. Images were pre-processed using the \textit{crop} method, and the model was trained for five epochs. This result is compared with NIIRS as computed with the GIQE5. The result is shown in Figure \ref{fig:cornerstone}. We choose to compare rAP in most experiments, because it can be generalized to unknown classes, as shown in Section \ref{subsec:generalize}. We refer to the baseline experiment in the following experiment subsections, in most cases with just one parameter altered from Table \ref{tab:experiment_parameters}.

If we return to the original formulation from Equation \ref{principle_equation}, we can consider the baseline problem as:
\begin{equation}
\argmax_{f}{\text{rAP}(f; P, m, \mathcal{D})}
\label{principle_equation_ex2}
\end{equation}
where $m$ is DenseNet161, $\mathcal{D}$ is the fMoW subset, and $P$ corresponds to all image system parameters except for $f$, the focal length. 

We must take care to discern the performance metrics exhibited here from NIIRS. While these experiments measure how well a classifier performs at recognizing specific objects, NIIRS measures how well all objects can be interpreted at each tier of its scale. For this reason, we focus on comparing optima between the two, instead of attempting to relate absolute performance. In this case, we compare:
\begin{equation}
\argmax_{f}{\text{rAP}} \;\; \text{vs.} \; \argmax_{f}{\text{NIIRS}}
\label{principle_equation_comparison}
\end{equation}

\begin{figure}[!t]
\centering
\includegraphics[trim={0.25cm 0 0 0},width=0.49\textwidth]{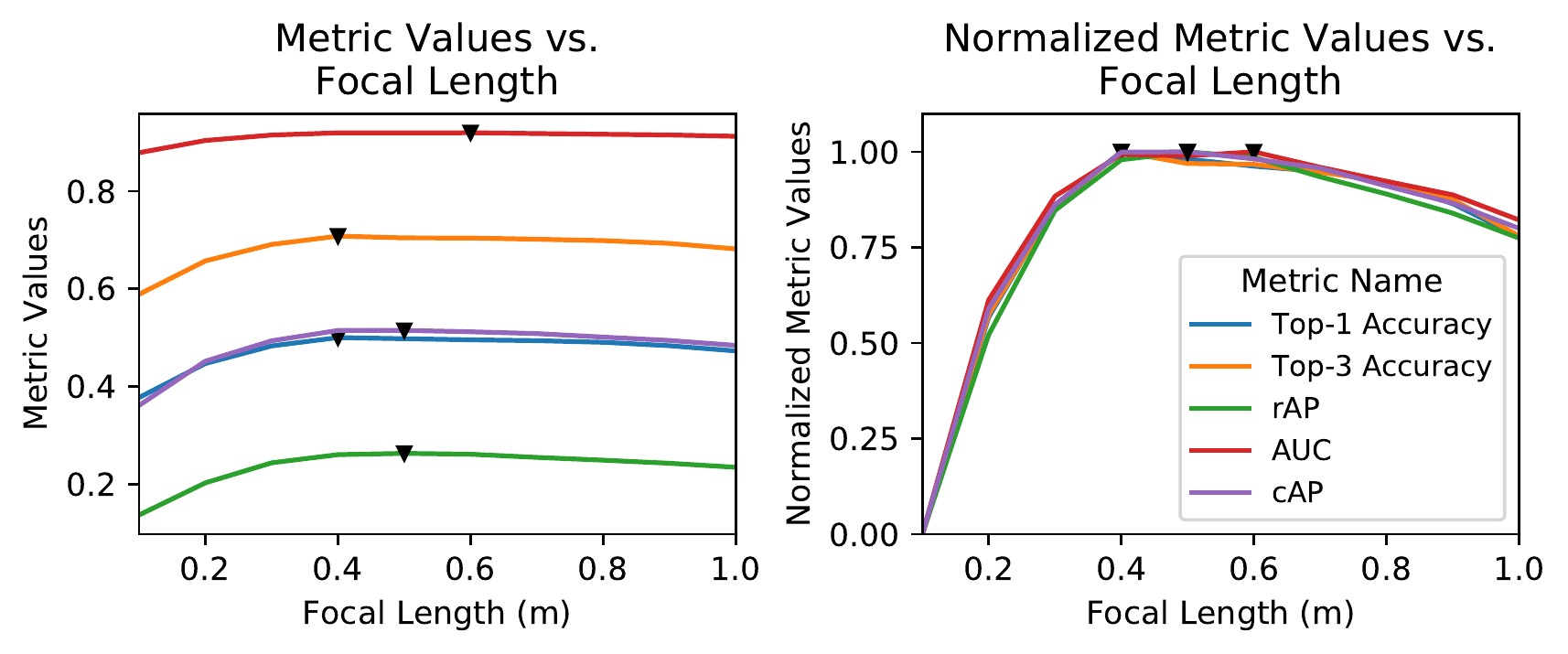} 
\caption{Multiple metrics are shown  for validation set performance of the baseline model as a function of varied focal length. Notice that the normalized curves are nearly identical, but peak at different focal length values. The importance of this variation will depend on precision and cost requirements for a given application.}
\label{fig:multi_stat}
\end{figure}

We show the performance relationship for the baseline experiment for all considered problems and metrics in Figure \ref{fig:multi_stat}. Notice that the two accuracy metrics peak at $f=0.4$, the two AP metrics peak at $f=0.5$, and the AUC metric peaks at $f=0.6$. Since all five metrics are near their peak from $f \in [0.4, 0.6]$, focal length may be selected based on another consideration within this range, such as cost.

\begin{figure*}[!t]
\centering
\includegraphics[width=\textwidth]{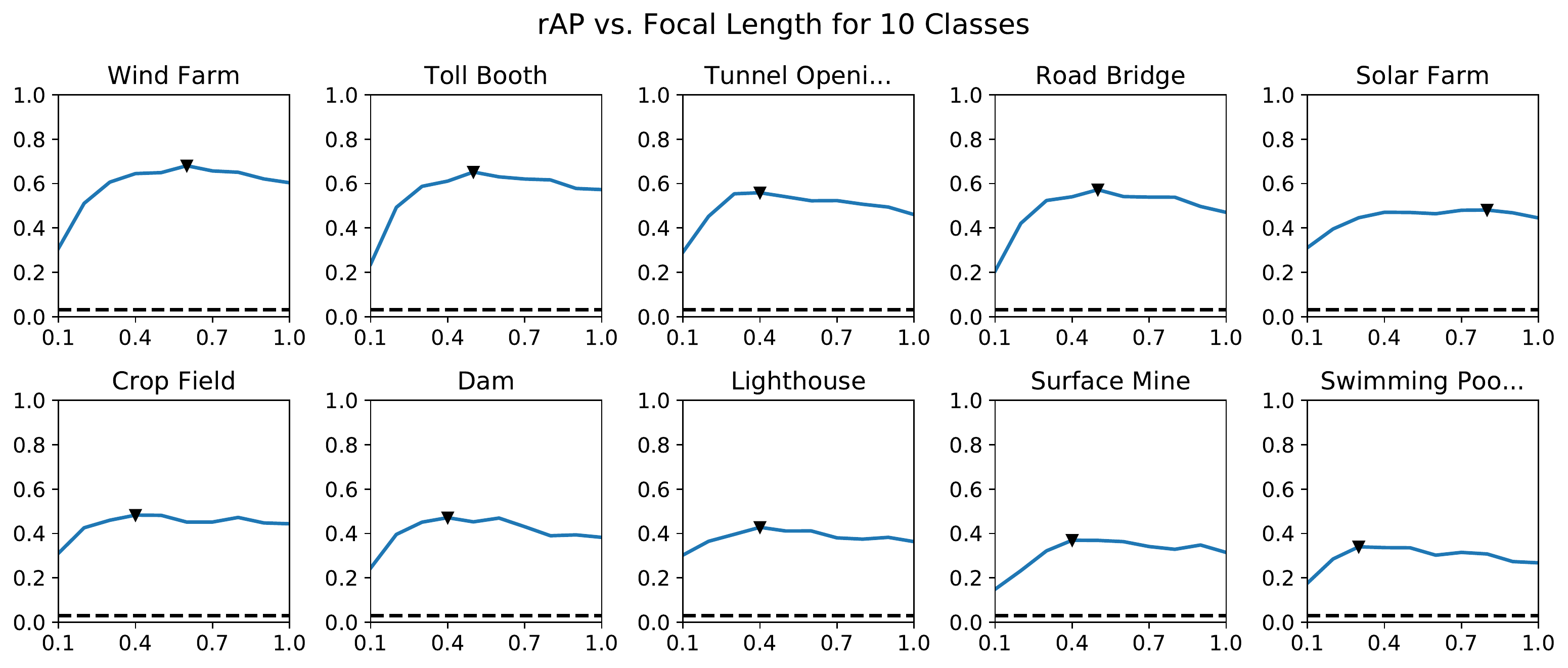} 
\caption{Plots of rAP are shown for 10 different classes for the baseline experiment, as opposed to averages used for the other experiments. The black dotted line in each plot indicates a lower bound for rAP, if the image embeddings were randomly distributed.}
\label{fig:per_class_plots}
\end{figure*}

We give a breakdown of rAP performance for ten individual classes in Figure \ref{fig:per_class_plots}. While the relationship between focal length and rAP for each class is similar, there is significant variation in the focal length value at which rAP is greatest, e.g., 0.3 for swimming pool vs. 0.8 for solar farm. This could be explained by reliance on spectral vs. spatial features. At lower focal length values, spectral features are better preserved than spatial features, explaining why the bright blue of the swimming pool would be more easily recognized. At higher focal length values, as less light enters the system and more noise is introduced, spatial features have greater weight than spectral features, explaining why the gridded pattern of the solar farm would be more easily recognized. In addition, some classes have a greater range in rAP with respect to focal length, e.g., 0.4 for road bridge vs. 0.2 for crop field. These observations indicate that the spectral and spatial features of the imagery in these classes, independent of the overall quality of the imagery, have a variable response with respect to changes in resolution and noise.

\subsection{Varying Aperture Diameter}
\label{subsec:diam}
\begin{figure*}[!t]
\centering
\includegraphics[width=\textwidth]{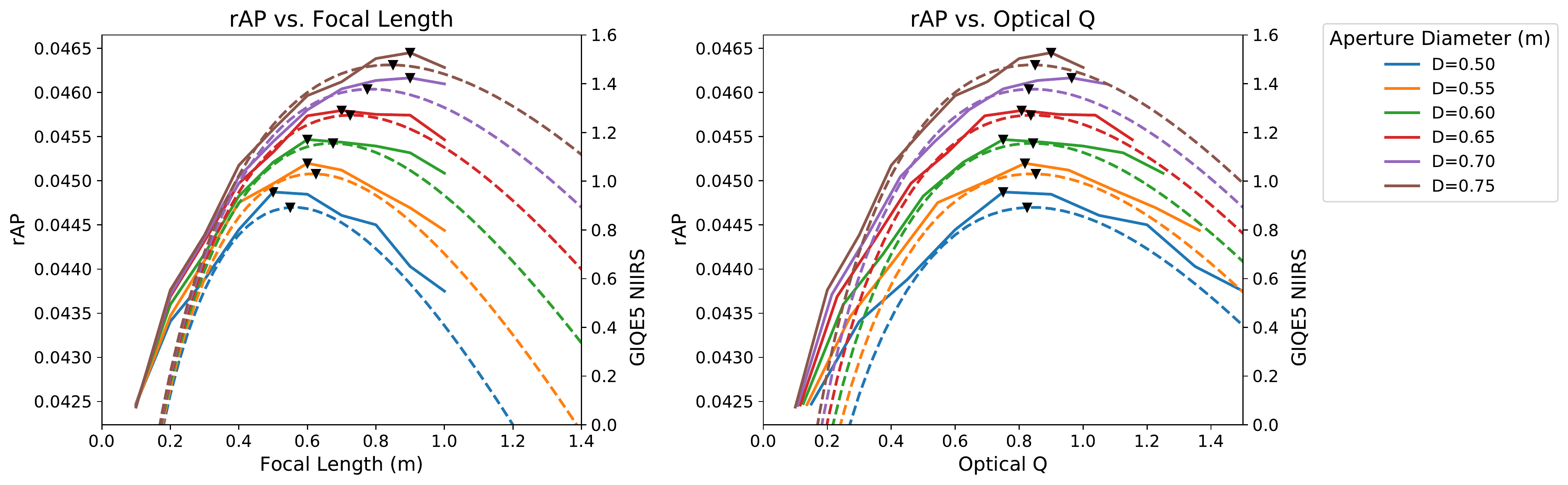} 
\caption{Baseline system evaluated directly from pre-trained weights, showing changing rAP as a function of focal length and Q for six aperture diameter values. The solid lines correspond to experimental results, and the dotted lines to GIQE5 NIIRS. Notice the similarity to the baseline result, despite the network not having seen any overhead images during training.}
\label{fig:vary_diameter_epoch0}
\end{figure*}

We here explore variation of optic aperture diameter, in addition to focal length. First, we consider how a model pre-trained on ImageNet, but without fine-tuning on the fMoW subset, will perform as aperture diameter changes. The results of this experiment are shown in Figure \ref{fig:vary_diameter_epoch0}. As mentioned in Section \ref{subsec:aftersim}, this was the only scenario in which standard score normalization (ImageNet mean, standard deviation) was used. 

\begin{figure*}[!t]
\centering
\includegraphics[width=\textwidth]{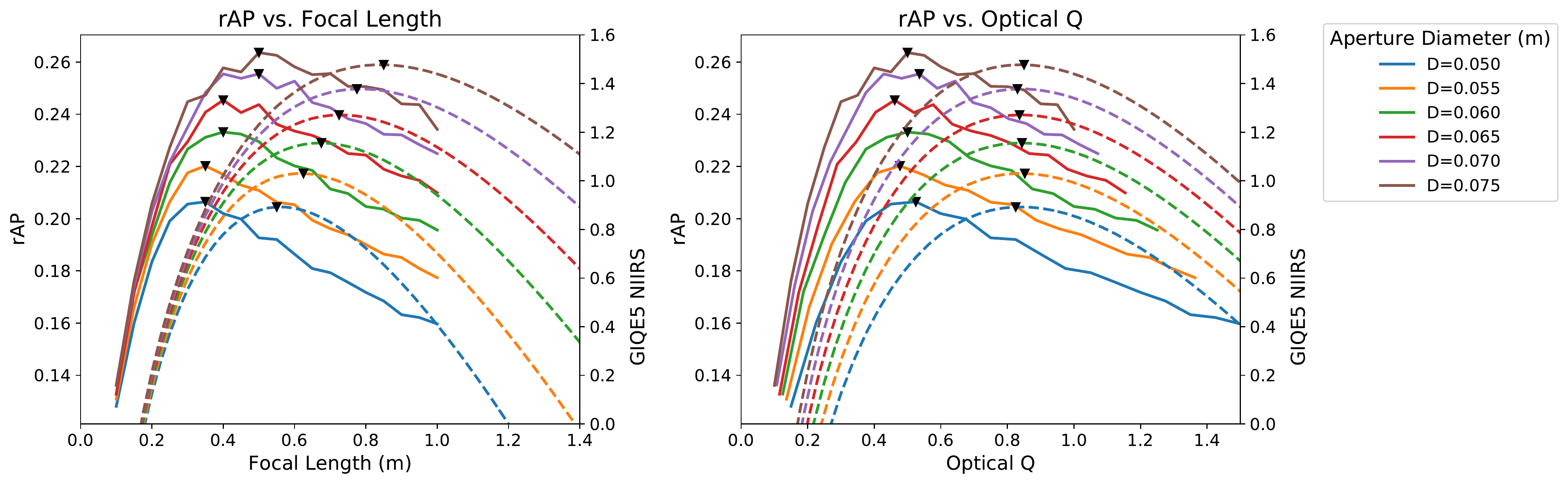} 
\caption{Baseline system fine-tuned from pre-trained weights, showing changing rAP as a function of focal length and Q for six aperture diameter values. The solid lines correspond to experimental results, and the dotted lines to GIQE5 NIIRS. Notice that the peak rAP values, denoted by triangles, are closely aligned in the plot vs. Q on the right, unlike in the plot vs. focal length on the left. This shows that Q is significant in describing image utility.}
\label{fig:vary_diameter_niirs}
\end{figure*}

Then, we consider how the pre-trained model will perform once it has been fine-tuned for five epochs, just as in the baseline. The results of this experiment are shown in Figure \ref{fig:vary_diameter_niirs}. First, notice that the relationship between the systems with different aperture diameter is similar for low focal length, but diverges as focal length increases. This can be understood through Figure \ref{fig:d_compare}, in which the quantity of noise present in the low-aperture diameter system becomes significantly greater as focal length increases.

Next, notice that the pre-trained model without fine-tuning better corresponds to GIQE5 NIIRS than the fine-tuned model. While the axes have been adjusted to better compare against NIIRS, notice that the rAP values on the left y-axis of Figure \ref{fig:vary_diameter_niirs} are much greater than in Figure \ref{fig:vary_diameter_epoch0}. Therefore, while the alignment between NIIRS and rAP in \ref{fig:vary_diameter_epoch0} may appear significant, a designer would more likely want to optimize for the relationship exhibited in  \ref{fig:vary_diameter_niirs}, in which the model performance is greater.

Still, this result raises an interesting question: could it be the case that the CNN pre-trained on ImageNet has performance more similar to the GIQE because it more closely models human recognition? In this case, improved performance when fine-tuned on the satellite imagery could suggest that when the network ``specialized'' towards a specific domain, it not only gained improved performance over a generic recognizer, but attained an altered capability with respect to tolerance of artifacts. From an evolutionary perspective, it is unreasonable to assume that the human visual system is perfectly attuned to this domain of imagery, and to optimally extract patterns when some combination of noise and blurring are present. Further, a computational solution may not be not phased by artifacts such as aliasing, which look displeasing to humans, but do not significantly alter the semantic information present.

\subsection{Varying Architectures}
As observed in comparing Tables \ref{tab:arch_info} and \ref{tab:orig_performance}, different architectures may be optimal for different datasets and problems. Therefore, we consider it important to compare varying architectures, to test if the same parameter-performance relationship manifests. 

\begin{figure}[!t]
\centering
\includegraphics[trim={0.25cm 0 0 0},width=0.49\textwidth,draft=false]{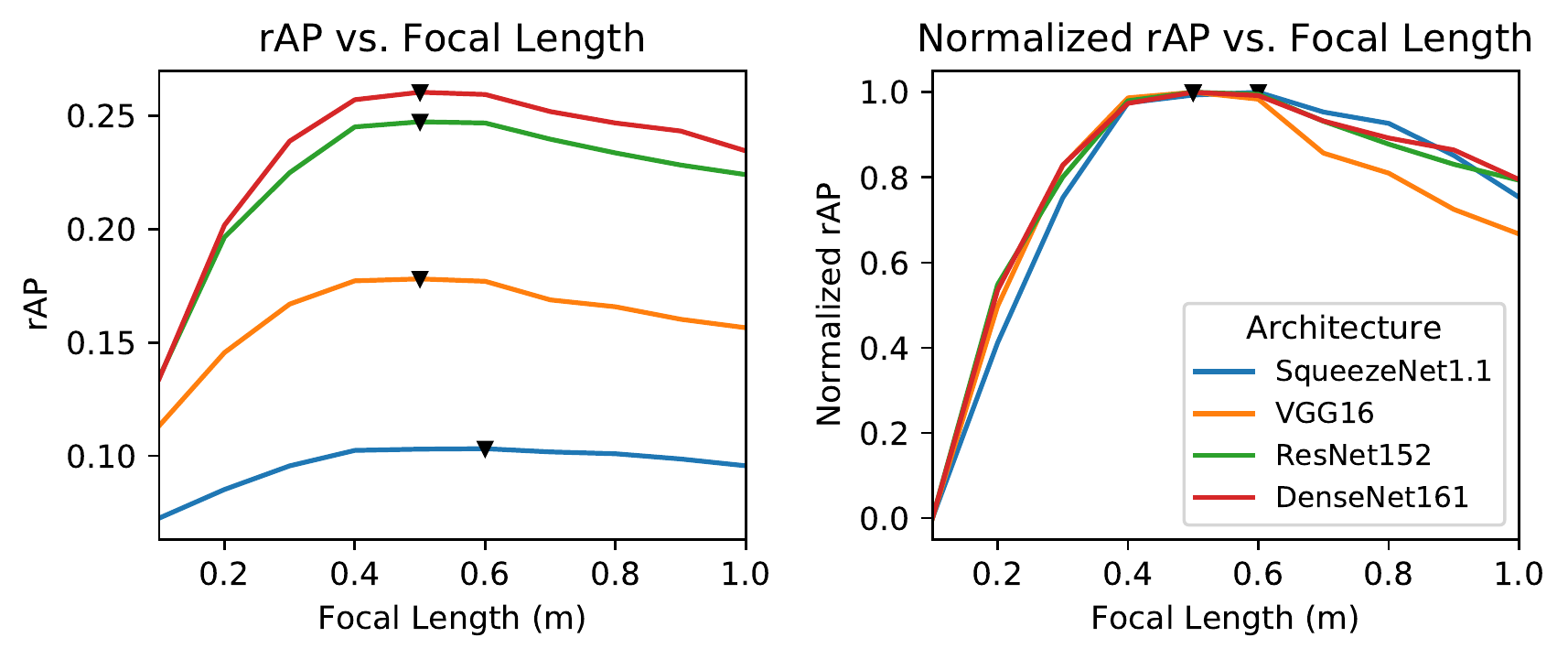} 
\caption{SqueezeNet1.1, VGG16, ResNet152, and DenseNet161 are compared for the baseline experiment. Notice that while SqueezeNet1.1 has a different optimal focal length of 0.6m vs. 0.5m for the other architectures, the normalized curves are very similar.}
\label{fig:vary_arch}
\end{figure}

In Figure \ref{fig:vary_arch}, the three architectures described in Section \ref{subsec:cnn} are compared for the baseline experiment. All three models attained peak performance at the same focal length value, suggesting that the models reacted similarly to variation in resolution and noise, even while they achieved different levels of absolute performance. This result indicates a consistency with respect to visual recognition with CNNs. This consistency, combined with the difference between CNN performance and GIQE5 NIIRS shown in Figure \ref{fig:vary_diameter_niirs}, could suggest that CNNs in general are differently affected by resolution and noise than the HVS.

\subsection{Varying Learning Objective}
\begin{figure}[!t]
\centering
\includegraphics[trim={0.25cm 0 0 0},width=0.49\textwidth]{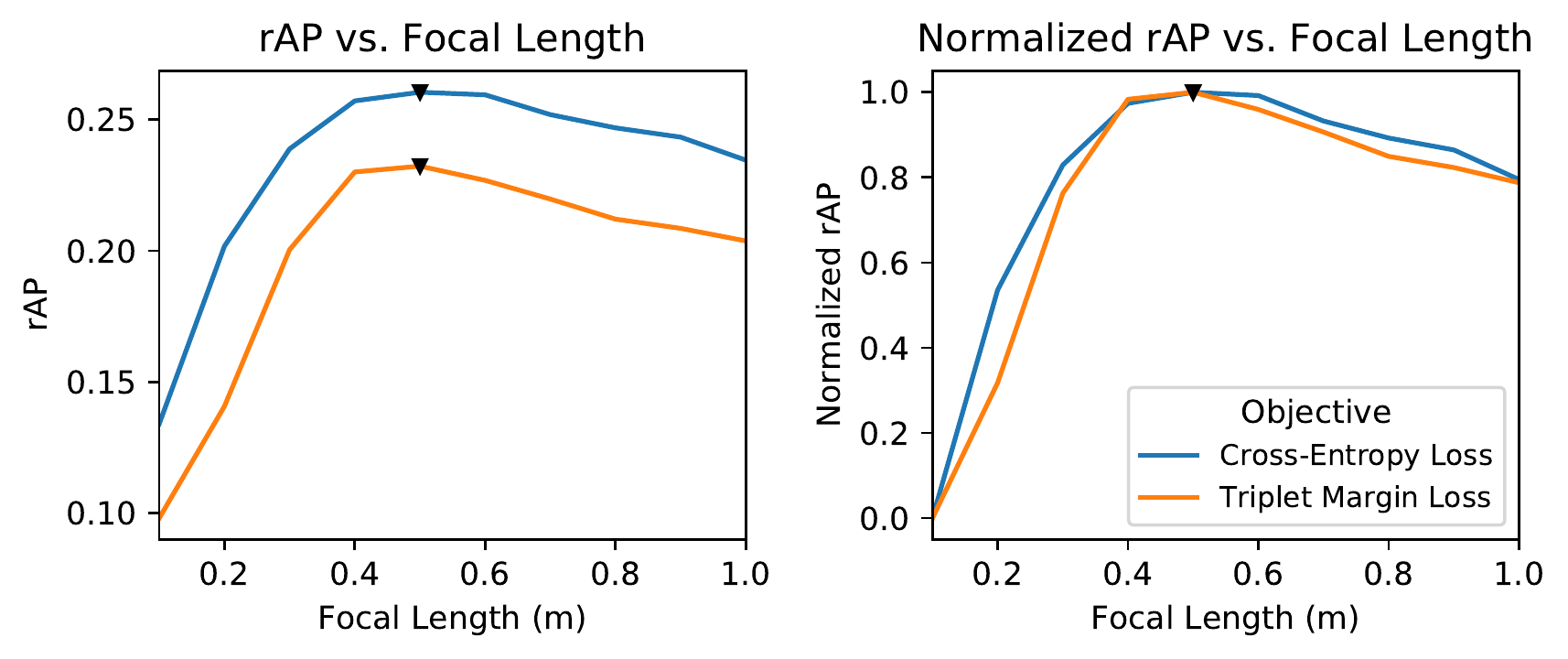} 
\caption{Cross-entropy loss training and triplet margin loss training are compared for the baseline experiment with DenseNet161. While triplet training does not perform as well as cross-entropy training after five epochs, triplet training likely requires more epochs to finish its optimization. Still, we note that the normalized performance curves are nearly identical.}
\label{fig:vary_triplet_arch}
\end{figure}

In this experiment, we compare the performance of the baseline scenario with cross-entropy loss training vs. triplet margin loss training. The goal is to test whether the same focal length-rAP relationship manifests when optimizing for two significantly different objectives. As described in Section \ref{subsec:mod_mod}, the two variations of DenseNet161 used differ in the last two layers. The results shown in Figure \ref{fig:vary_triplet_arch} indicate that the focal length-rAP relationship is consistent between the two objectives. 

\subsection{Varying Classes}
While experimentation on additional overhead image datasets such as SpaceNet and xView would be ideal, fMoW contains such a breadth of imagery that it serves as a strong proxy for any recent multispectral overhead imagery. To account for bias in observing all classes of the dataset simultaneously, we have designed an experiment in which only disjoint subsets of the class space are evaluated.

\begin{figure}[!t]
\centering
\includegraphics[trim={0.25cm 0 0 0},width=0.49\textwidth]{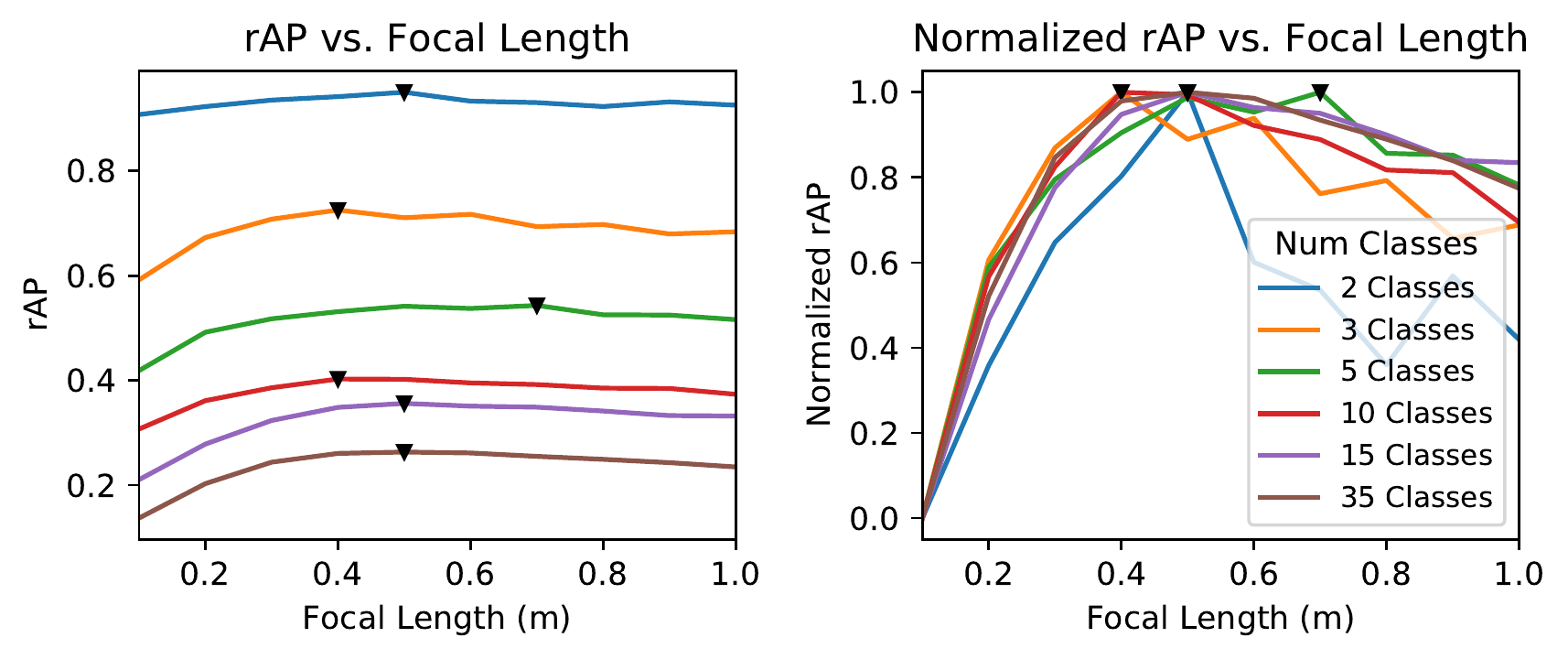} 
\caption{The fMoW subset is partitioned into 5 sets disjoint by class, containing 2, 3, 5, 10, and 15 of the total classes respectively. The baseline experiment is conducted on each of these subsets and results are compared against the baseline with all 35 classes.}
\label{fig:vary_classes}
\end{figure}

The results of this experiment are shown in Figure \ref{fig:vary_classes}. As anticipated from Figure \ref{fig:per_class_plots}, partitions of the fMoW subset containing few classes have a more erratic relationship. This result reinforces the point that the classes observed are important to the sensor design problem. We recommend that several different classes are observed in order to make robust sensor design decisions. Figure \ref{fig:vary_classes} indicates that the focal length-rAP relationship becomes stable once around 15 classes are used, for this experiment.

\subsection{Varying Epochs Trained}
\begin{figure}[!t]
\centering
\includegraphics[width=0.5\textwidth]{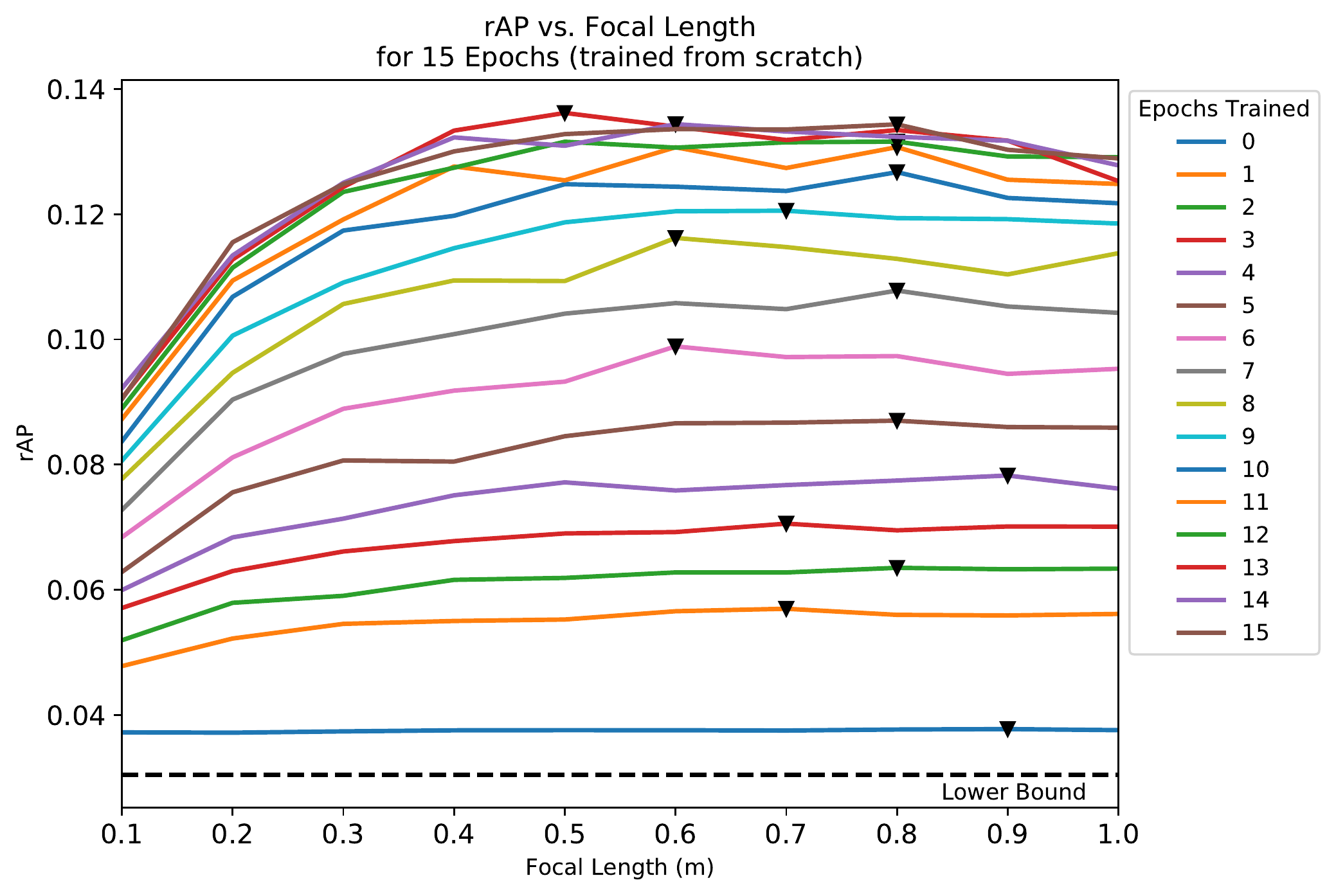} 
\caption{The baseline system is trained from scratch, showing changing rAP as a function of focal length for each training epoch. The peak rAP across focal length values for a given epoch is denoted with a triangle. Notice that the peak value shifts towards a smaller focal length as the model is trained for more epochs.}
\label{fig:per_epoch_scratch}
\end{figure}

\begin{figure}[!t]
\centering
\includegraphics[width=0.5\textwidth]{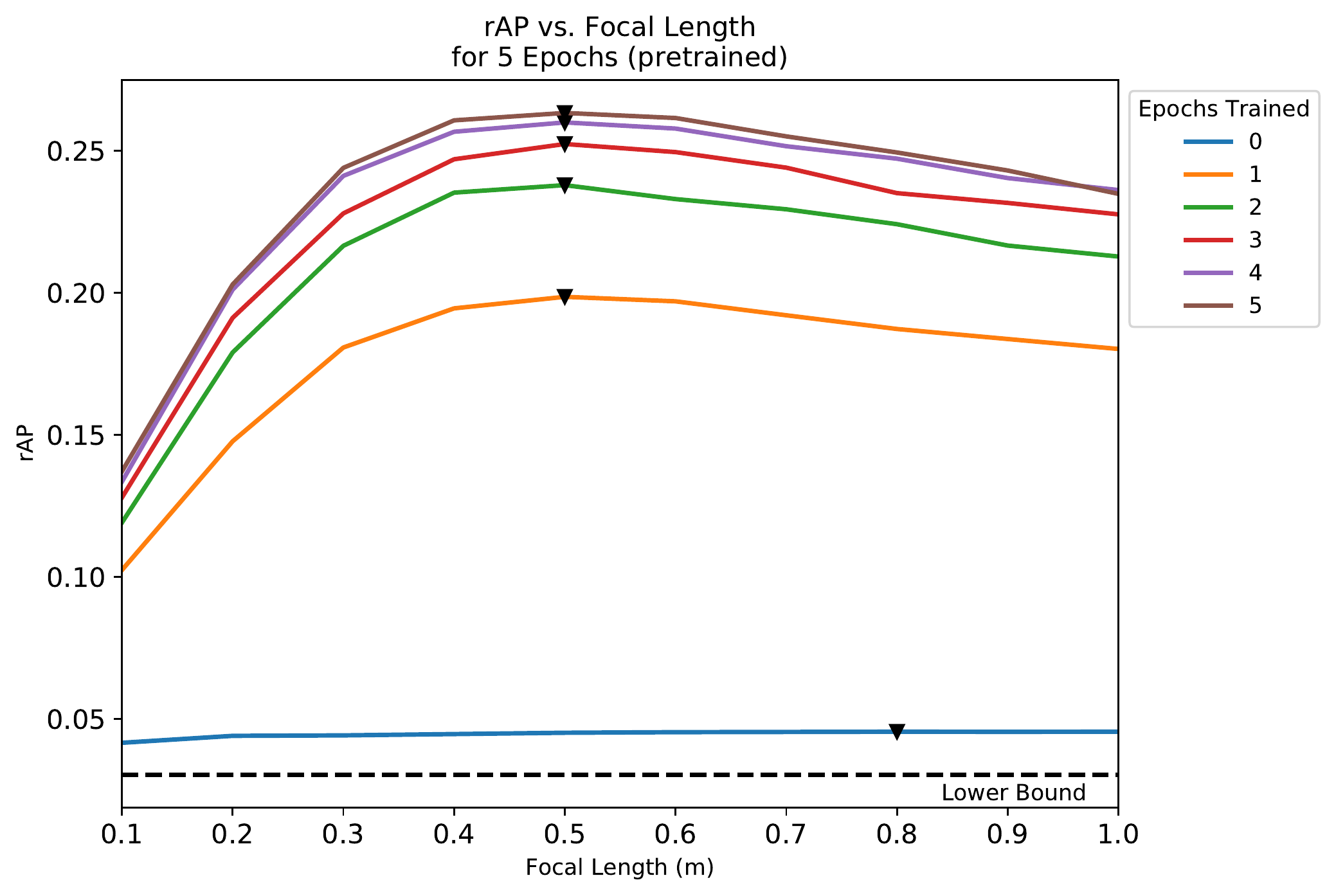} 
\caption{The baseline system is trained from pre-trained weights, showing changing rAP as a function of focal length for each training epoch. The peak rAP across focal length values for a given epoch is denoted with a triangle. Notice that the peak value remains consistent as the network is trained.}
\label{fig:per_epoch_pretrained}
\end{figure}

Here, we consider how many epochs a model must be trained for in order to converge on a stable focal length-rAP relationship. We consider both training from scratch and fine-tuning of a pre-trained model in Figures \ref{fig:per_epoch_scratch} and \ref{fig:per_epoch_pretrained} respectively. In the from-scratch scenario, we note not only that many more epochs are required to reach a stable relationship, but also that absolute performance is considerably reduced. In the fine-tuning scenario, one training epoch alone may be sufficient to exhibit a stable relationship.

\subsection{Varying Pre-Processing}
As discussed in Section \ref{subsec:preproc}, we considered two different methods of pre-processing the image data, in order to fit it into the fixed-size format required for the CNN models. We note that most top competitors in the fMoW challenge used a method similar to resizing, but with additional ``context'' pixels around the target bounding box from the original image. While this greatly improves performance for small targets, we wanted to avoid adding additional heuristics to the data preparation process. 

\begin{figure}[!t]
\centering
\includegraphics[trim={0.25cm 0 0 0},width=0.49\textwidth]{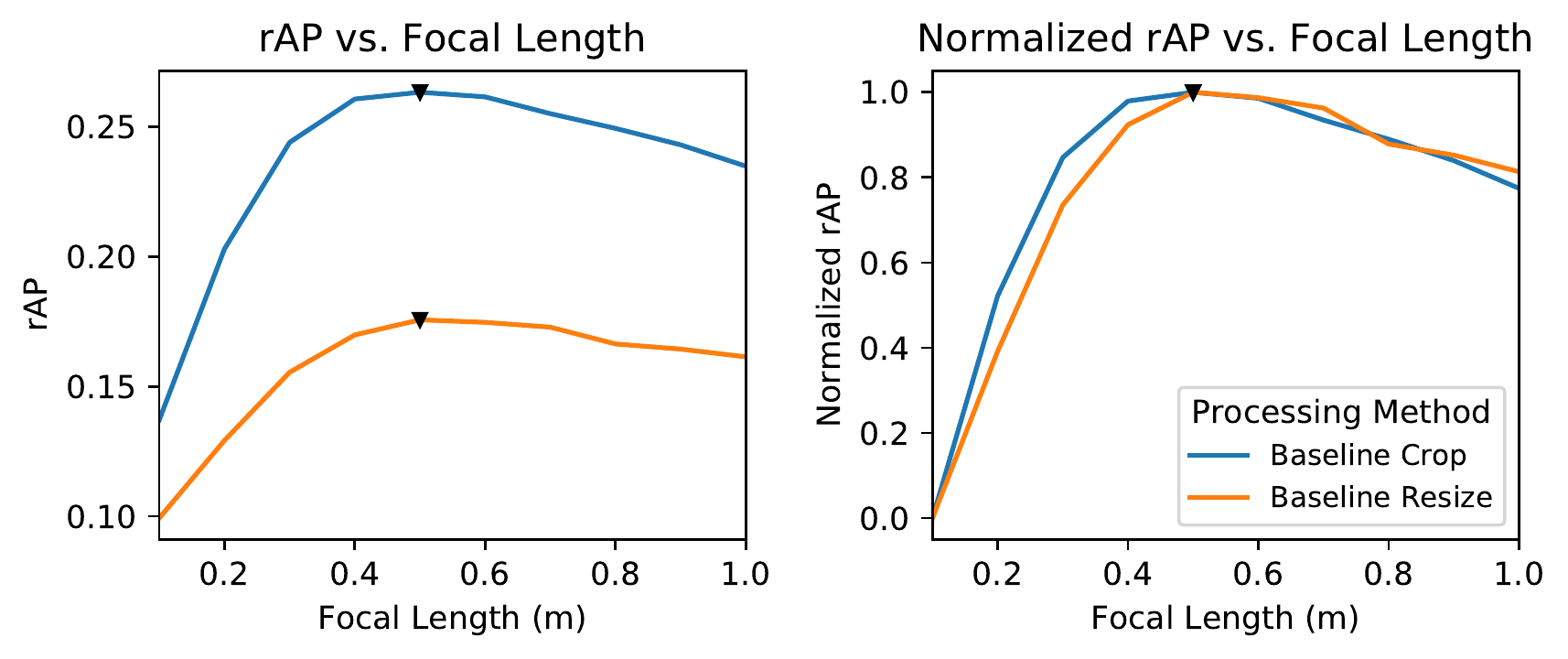} 
\caption{The baseline system is trained with images cropped vs. images resized. Since the images have such a large variation in size, as shown in Figure \ref{fig:fmow_area}, the pre-processing method used for input into the fixed-size network is critical.}
\label{fig:var_proc}
\end{figure}

A comparison of performance for the \textit{cropping} and \textit{resizing} methods is shown in Figure \ref{fig:var_proc}. While there is a significant gap in absolute performance, as shown in the plot on the left, the normalized curves on the right are quite similar. This further fortifies our conclusion that there is a self-consistency to visual recognition with CNNs.

\subsection{Zero-Shot Experiment}
\label{subsec:generalize}
\begin{figure}[!t]
\centering
\includegraphics[trim={0.25cm 0 0 0},width=0.49\textwidth,draft=false]{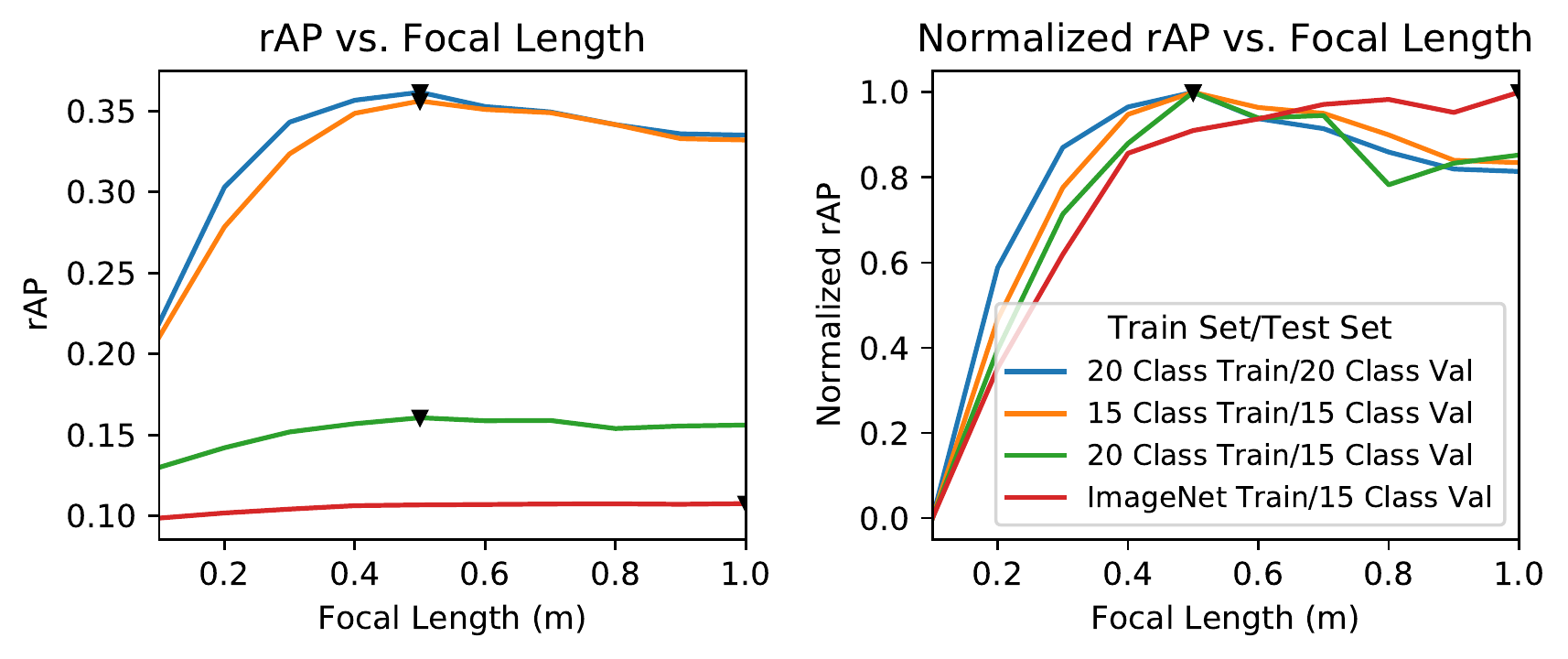} 
\caption{This plot demonstrates the zero-shot experiment for the baseline case. Four different relationships are shown in this plot: First, the 20-class and 15-class partitions are considered separately. Each was trained and validated on without observing the other. Then, the trained models from the 20-class partition were used to evaluate the 15-class partition- this is the zero-shot case. Notice that these three relationships are very similar and have the same peak for the normalized rAP shown on the right. In addition, performance of the pre-trained ImageNet features on the 15-class set is demonstrated. The result is a relationship clearly unlike the others, indicating that these features may not be salient for this problem.}
\label{fig:generalize}
\end{figure}

\begin{figure}[!t]
\centering
\includegraphics[trim={0.25cm 0 0 0},width=0.49\textwidth,draft=false]{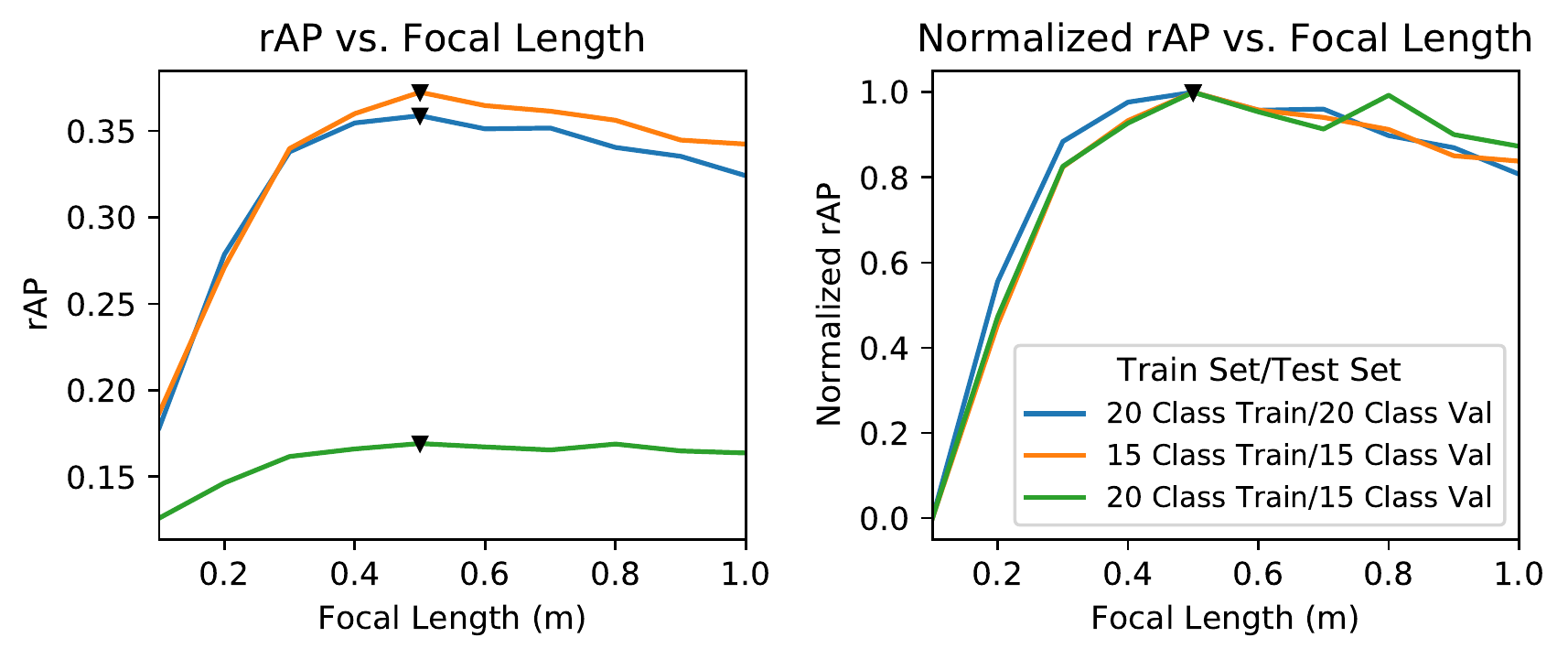} 
\caption{This plot mirrors Figure \ref{fig:generalize} for triplet training instead of classifier training. The normalized relationship is nearly identical, but the zero-shot rAP performance (20 Class Train/15 Class Val) is improved over the classifier case. This is logical, because the features learned with the triplet margin objective are more likely to apply to unseen classes, since the objective doesn't encode for any specific class.}
\label{fig:generalize_triplet}
\end{figure}

As explained in Section \ref{subsec:experiments}, in this experiment, the fMoW subset is partitioned into two sets disjoint by class, with the training set containing 20 classes, and the validation set containing 15 classes. The model trained on the training set never sees any examples from classes in the validation set. The results are shown in Figures \ref{fig:generalize} and \ref{fig:generalize_triplet} for classifier and triplet training respectively.

This is perhaps the most important experiment conducted in this work, as it shows that the model trained on 20 fMoW classes generalizes to the remaining 15. This suggests that the model could apply to data annotated from any DigitalGlobe sensor, and perhaps from other remote sensors as well. We posit that this fine-tuned model can be used to conduct the sensor parameter space analysis for a wide variety of useful scenarios.

\subsection{IQA}
\begin{figure}[!t]
\centering
\includegraphics[width=0.5\textwidth,draft=false]{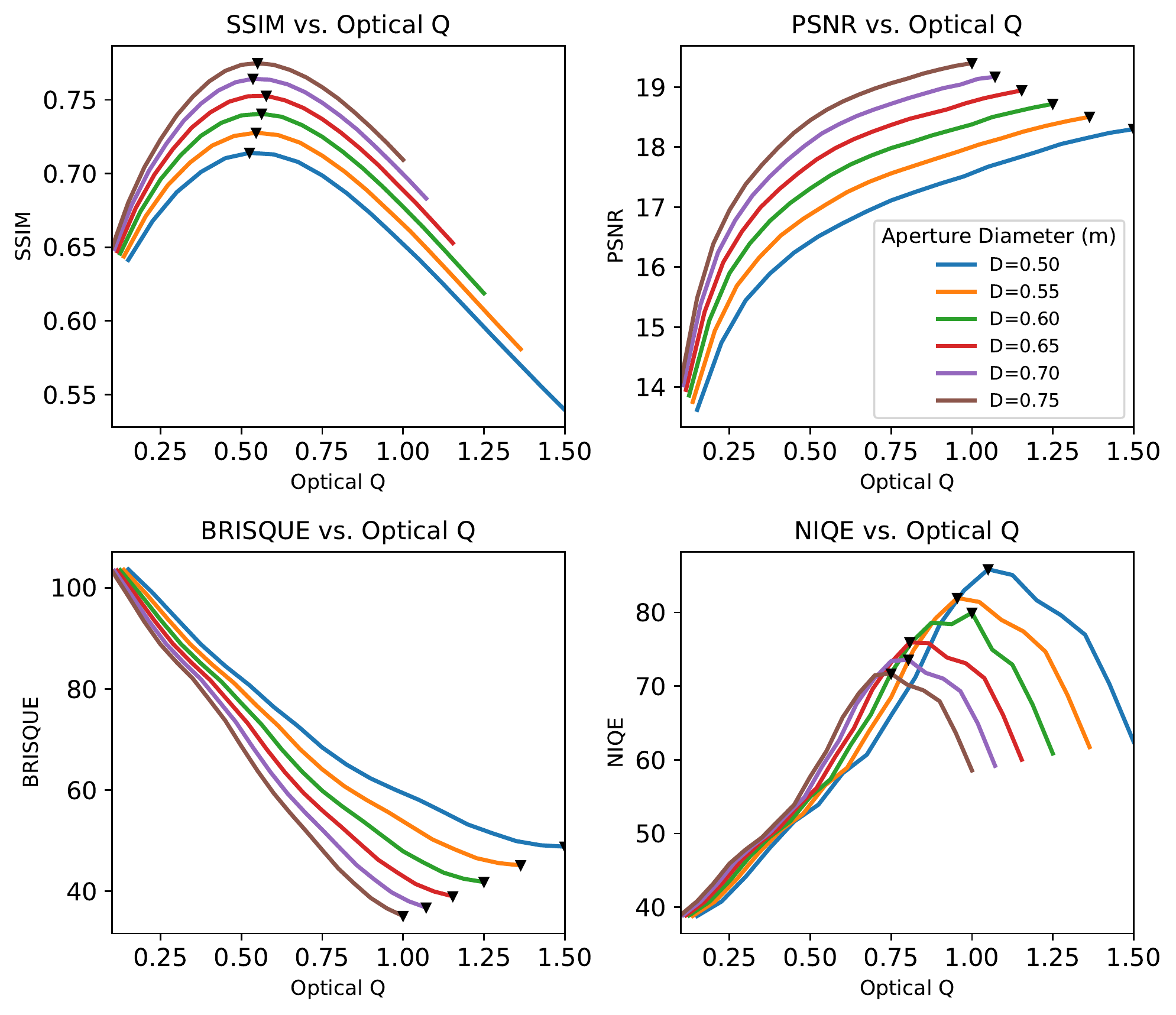} 
\caption{In this plot, we repeat the experiment from Section \ref{subsec:diam} with four different IQA metrics. Six different aperture diameter values are tested, in addition to the varying focal length, and the results are plotted against optical Q. For the full-reference metrics, PSNR increased monotonically, while SSIM has a clear optimal aperture diameter at around Q=0.5. For the no-reference metrics, BRISQUE decreases montonically, while NIQE has an optimal point for each trial, but those optima do not correspond to any specific Q value.}
\label{fig:iqa}
\end{figure}

Finally, we compare the performance of the various IQA algorithms for the baseline parameter configuration. The results of this experiment are shown in Figure \ref{fig:iqa}. While both PSNR and BRISQUE have monotonic relationships with respect to increasing focal length, SSIM and NIQE have clear optima in the tested range. Further, the SSIM plot exhibits an optimal Q value, similar to the relationship we saw for CNN performance and NIIRS. 

In Table \ref{tab:optimal_q}, the results from four experimental trials, that varied both focal length and aperture diameter, are summarized to show the Q value for which each metric was optimal. The mean and standard deviation of this Q value are shown, in addition the the minimum and maximum values of the metric measured at this Q value. This result shows that the optimal Q for pre-trained CNN performance is closer to SSIM than to the other metrics. This indicates that out of the four IQA metrics tested, only SSIM may be useful for understanding sensor system performance.

\begin{table}
\centering
\captionof{table}{Optimal Q for Different Metrics} \label{tab:optimal_q} 
\begin{tabular*}{0.48\textwidth}{l @{\extracolsep{\fill}} rrrr}
\toprule[1.5pt]
\textbf{Metric} & $\mu(\text{Q})$ & $\sigma(\text{Q})$  & val$_{min}$ & val$_{max}$ \\
\midrule
Scratch rAP* &  0.834 & 0.065 & 0.045 & 0.046 \\
Pre-Trained rAP* &  0.481 & 0.029 & 0.206 & 0.263 \\
GIQE5 NIIRS & 0.840 & 0.010 & 0.892 & 1.477 \\
SSIM & 0.549 & 0.017 & 0.714 & 0.775 \\
\bottomrule
\end{tabular*}
\begin{tablenotes}
\item[*]
This table shows that certain metrics exhibit an optimal value for optical Q when both focal length and aperture diameter were varied.

*Scratch rAP and Pre-Trained rAP refer to the values from Figures \ref{fig:vary_diameter_epoch0} and \ref{fig:vary_diameter_niirs} respectively. 
\end{tablenotes}
\end{table}

%% file: conclusion.tex
\section{Conclusion}
\label{sec:conclusion}
In this work, we develop a methodology for optimizing remote sensing image system parameters with respect to performance of deep learning models on real overhead image data. We demonstrate a tool which implements this methodology, and conduct a variety of experiments using different sensor and learning parameters.

Through these experiments, we demonstrate that visual recognition performance for CNNs is self-consistent under a variety of conditions. We also show that human and machine visual recognition performance may differ significantly. Specifically, we show that the most recent version of the GIQE does not correspond to CNN performance, within the observed parameter space. It is still possible that CNN performance is closer to true NIIRS than observed, in the event that the GIQE does not accurately compute NIIRS in this parameter space.

We show that the SSIM full-reference IQA algorithm produces similar results to CNN performance. If SSIM could be modified to match CNN performance (across a more expansive sensor parameter space), this would circumvent the need to conduct costly CNN experiments for sensor system optimization. Further, this is a step in the direction of modifying the GIQE itself to reflect this relationship, circumventing the need for a data-based optimization altogether. Still, these surrogate approaches can only reveal relative optima for sensor parameters. In order to understand absolute model performance for some data, applying our method with the target model and data will produce the most reliable result.

An implication of this work is that all overhead systems which have been designed to optimize for NIIRS using the GIQE, and which produce imagery which is analyzed with CNNs, have been designed non-optimally. Designing systems based on the methods described in this paper could yield more optimal systems for CNN processing. Further, these methods can be applied to any visual recognition model, as shown through the generalization posed by Equation \ref{principle_equation}. When a visual recognition model superior to CNNs is introduced, the same study can be conducted and compared against the results presented here.

This concept can be extended to other domains, including sensing for autonomous cars and machine vision in an industrial setting. We posit that optimizing sensors for visual recognition problems will become an important part of designing automated sensing systems.

In the future, this work will be expanded to study different datasets, apply more complex CNN models, and develop IQA and GIQE models which match the observed CNN parameter-performance relationship. 

%% file: postamble.tex
\section*{Acknowledgments}
The authors would like to thank Daniel LeMaster from AFRL for providing access to his pyBSM code and helpfully answering our questions.

\ifCLASSOPTIONcaptionsoff
  \newpage
\fi

\bibliography{references}
\bibliographystyle{ieeetr}